\documentclass[prb,twocolumn,showpacs, showkeys,amsmath,amssymb,times]{revtex4-1}

\usepackage{graphicx}
\usepackage{epstopdf} 
\usepackage{dcolumn}
\usepackage{bm}
\usepackage{hyperref}
\usepackage[mathlines]{lineno}
\usepackage[usenames,dvipsnames,svgnames,table]{xcolor}
\usepackage{times}

\newcommand{\beq}{\begin{equation}}
\newcommand{\eeq}{\end{equation}}
\newcommand{\beqn}{\begin{eqnarray}}
\newcommand{\eeqn}{\end{eqnarray}}
\newcommand{\he}[1]{\textsuperscript{#1}He}

\begin{document}
\title{Ultrasonic Interferometer for First-Sound Measurements of Confined Liquid $^4$He}

\author{X. Rojas}
\email{xavier@ualberta.ca}
\author{B. D. Hauer, A. J. R. MacDonald, P. Saberi, Y. Yang}
\author{J. P. Davis}
\email{jdavis@ualberta.ca}
\affiliation{Department of Physics, University of Alberta, Edmonton, Alberta, Canada T6G 2E9}
\date{\today}

\begin{abstract}
We present a new technique for probing the properties of quantum fluids in restricted geometries. We have confined liquid $^4$He within microfluidic devices formed from glass wafers, in which one dimension is on the micro scale. Using an ultrasonic analog to Fabry-P\'erot interferometry, we have measured the first-sound of the confined liquid $^4$He, which can be a probe of critical behavior near the lambda point ($T_{\lambda}$). All thermodynamic properties of liquid \he4 can be derived from first-sound and heat capacity measurements, and although quite a bit of experimental work has been done on the latter, no measurement of first-sound has been reported for a precisely confined geometry smaller than a few tens of micrometers. In this work, we report measurements of isobaric first-sound in liquid \he4 confined in cavities as small as $\sim5$ $\mu$m. Our experimental setup allows us to pressurize the liquid up to $\sim25$ bar without causing deformation of the confined geometry, a pressure which is about four times larger than previously reported with similar microfluidic devices. Our preliminary results indicate that one can possibly observe finite-size effects and verify scaling laws, by using similar devices with smaller confinement.
\end{abstract}



\maketitle


\section{\label{sec:level1_int}Introduction}
The confinement of quantum liquids in precisely defined geometries opens the door to new opportunities~\cite{Duh2012}, providing a laboratory for the study of \he3 topological superfluids~\cite{Levitin2013}, for which recent theoretical predictions show the emergence of new exotic superfluid phases~\cite{Vorontsov2007, Sauls2011,Wu2013, Wiman2013} in confinement. It also offers a framework for the study of finite-size effects in liquid \he4 at the superfluid transition~\cite{Garcia1999}, which provides the most precise test of scaling laws in critical phenomena~\cite{Gasparini2008} to date and allows the study of superfluid coupling and proximity effects~\cite{Perron2010, Perron2012}. However, in order to probe the properties of such a small amount of liquid (micro/nanoliters), new measurement techniques are necessary. Examples are the ultra-sensitive nuclear magnetic resonance (NMR) and calorimetry techniques developed in the Saunders~\cite{Levitin2007} and Gasparini groups~\cite{Mehta1998}, respectively. In this work, we present a new ultrasonic Fabry-P\'erot interferometry technique that we have used to probe the first-sound of confined liquid \he4, with the aim to explore finite-size effects near the superfluid transition. It is interesting to note that confinement geometry similar to ours could also be used as an apparatus for the study of superfluid resonators~\cite{DeLorenzo2013}, which have recently demonstrated quality factors as high as $7\times10^6$, or a platform for the coupling of superfluids to MEMS~\cite{Gonzalez2013,Gonzalez2013b}.

Near a second-order phase transition, such as the superfluid transition of \he4, a system moves into a critical regime characterized by locally correlated regions of the order parameter, which grow as the system approaches the transition point. At that critical point, the size of correlated regions, defined by the correlation length $\xi$, diverges.  The critical behavior of the correlation length can be studied via scaling laws, by measuring the thermodynamic response functions (specific heat, density, compressibility, etc.) that show singularities at the transition. In the case of the \he4 superfluid transition, the temperature dependence of the correlation length near the critical point is
\beq
\xi(t)=\xi_0 |t|^{-\nu},
\eeq
with $t=(T-T_\lambda)/T_\lambda$ the reduced temperature and $T_\lambda$ the transition temperature. The zero-temperature value of the correlation length, $\xi_0$, is on the order of the interatomic distance. The critical exponent $\nu$ dictates the critical behavior of the system and has been measured by many authors with very good precision, an example of which is the value $\nu=0.6705\pm0.0006$ obtained from second sound measurements performed by Goldner \textit{et al.}~\cite{Goldner1992}.

Thermodynamic singularities can be observed in a vast number of diverse physical systems which can be classified into a small number of universality classes. These universality classes are defined by the number of components of the order parameter and the dimensionality of the system. In the case of liquid \he4, the order parameter of the superfluid transition is a wavefunction characterized by two components (amplitude and phase) and so this system, along with many other systems (\textit{e.g.}, magnetic ordering transitions), belongs to the XY universality class. However, the superfluid transition of liquid \he4 presents unique advantages for the study of critical systems. Liquid \he4 is extremely pure and its only contaminant, \he3, can be removed by heat flushing techniques~\cite{Hendry1987} to as low as 1 part in $10^{15}$. Additionally, the order parameter of the superfluid transition is a wavefunction, and therefore not drastically influenced by the presence of the walls of the container~\cite{Mooney2002}. Moreover, the superfluid transition of liquid \he4 arises on a line of the pressure-temperature $(P$-$T)$ phase diagram, allowing its study over a range of pressures and temperatures.

As critical properties are related to the divergence of the correlation length at the critical point, one might be concerned about achieving a true thermodynamic limit (\textit{i.e.}, for a system of infinite size). Indeed, real systems are always finite in size so the correlation length can not diverge completely and will be limited to the size of the system. As a consequence, any thermodynamic response that is singular in the thermodynamic limit will no longer be so in a finite system, but will be rounded and exhibit a finite maximum as the transition region is approached. The cornerstone of our current understanding of the way thermodynamic singularities are modified by finite-size effects is the finite-size scaling theory introduced by Fisher~\cite{Fisher1972} in the early 70's. Experimentally, finite-size effects can be observed in several realizations, even when the confinement is not uniform (\textit{e.g.}\ superfluid transition of liquid \he4 confined in porous Vycor glass~\cite{Lambert1980}). It is, however, nearly impossible to verify finite-size scaling theories with such systems. The possibility of testing these theories depends crucially on available data measured under uniform confinement.

Liquid \he4 is an excellent candidate for testing the finite-size scaling theory of critical phenomena.  In the bulk limit, the properties of the superfluid transition of liquid \he4 are very well known~\cite{Goldner1992,Gasparini2008,Donnelly1998,Wilks1967}, for example, the data on the specific heat collected in micro-gravity by Lipa \textit{et al.}~\cite{Lipa2003} represents one of the most precise tests of theoretical predictions for critical phenomena. In addition, liquid \he4 wets almost all surfaces~\cite{Nacher1991} and it is therefore able to fill nano/microfluidic cavities designed for the study of finite-size effects. 

Recent work on finite-size effects in the specific heat at the liquid \he4 superfluid transition, by the Gasparini group, has shown remarkable agreement with finite-size scaling theories but have also exposed failures that remain unexplained~\cite{Gasparini2008}. The measurement of other thermodynamic responses, like first-sound, can provide new data to help resolve such problems. Additionally, the study of sound in liquid \he4 is particularly interesting because it is possible to use sound measurements (first, second, and fourth) to self-consistently determine all thermodynamic properties~\cite{Maynard1976,Maynard1980}.  Alternatively, the combination of first-sound and specific heat can be used to determine the full set of thermodynamic quantities~\cite{Maynard1976,Maynard1980}.

Near the transition, both the velocity and the attenuation of the first-sound exhibit critical behavior~\cite{Barmatz1968, Lambert1979, Ferrell1980}. In order to describe this critical behavior, a phenomenological theory has been developed by Ferrell and Battacharjee~\cite{Ferrell1980,Ferrell1987}, which can be justified to some extent within renormalization group theories~\cite{Pankert1989a,Pankert1989b}. The first-sound critical behavior of the bulk system is relatively well understood, nevertheless, we are unaware of any experiment performed in restricted geometries for which it has been possible to verify the finite-size scaling theories. Lambert~\textit{et al.}~\cite{Lambert1980} have measured the first-sound critical attenuation of liquid \he4 confined in porous Vycor glass, however, this porous material provides a disordered confinement which makes the interpretation of their results with finite-size scaling theories difficult. Other experiments~\cite{Lambert1979b, Lea1989} using ultrasound to study finite-size effects have been performed on thin \he4 films that have a free surface. However, this free surface leads to a film thinning~\cite{Garcia1999} as $T_{\lambda}$ is approached and make the interpretation of the results in the vicinity of the lambda point difficult. A precisely defined and homogeneous confinement, like the one provided by micro/nanofluidic devices, can help to verify finite-size scaling theories. Additionally, other properties of critical systems have been predicted for the superfluid transition of liquid \he4 in such homogeneous confinement. For example, Bhattacharyya~\textit{et al.}~\cite{Bhattacharyya1998} predict the emergence of different critical regimes for the ultrasonic attenuation according to the ratio of the confinement length to the acoustic path length, which is a function of the ultrasonic frequency.

In our experiment, we have confined liquid \he4 in the resonant cavity of an ultrasonic Fabry-P\'erot interferometer (UFPI). The confinement length (\textit{i.e.}\ the resonant cavity thickness) was made as small as $d=5.18\thinspace\mu$m, and its uniformity over the entire area has been precisely characterized by optical interferometry, resulting in a standard deviation for the cavity thickness of $\sigma=0.10\thinspace\mu$m. Our experimental setup has successfully allowed us to measure the first-sound velocity up to high pressures ($\sim24.8$ bar), without deformation of the cavity. This represents a noticeable improvement, as the pressure is about four times higher than reported in any other quantum microfluidic experiment~\cite{Levitin2013,Gasparini2008}. Our preliminary results do not show finite-size effects yet as the temperature resolution ($\sim50$ $\mu$K) is not sufficient for micron scale confinement~\cite{Gasparini2008}. However, our results agree well with known values of sound velocity in the bulk limit. The study of finite-size effects with this technique is possible with smaller confinement lengths.

\section{\label{sec:level1_exp}Experiment}
In this section, we describe the experimental setup, starting with the fabrication process of our microfluidic devices, which provide a thin circular cavity to confine liquid \he4 in a slab geometry. We give details on the characterization of the cavity thickness profile, which is a very important step for studying finite-size effects. Finally, we present the theory of the ultrasonic Fabry-P\'erot interferometer.

\subsection{\label{sec:level2_Experimental_setup}Experimental setup}
In order to fabricate a precisely defined confinement geometry we have used clean room nanofabrication techniques. The process is described briefly here but more technical details can be found in ref. [1].

\begin{figure}[t]
\centering
\includegraphics[width=8cm]{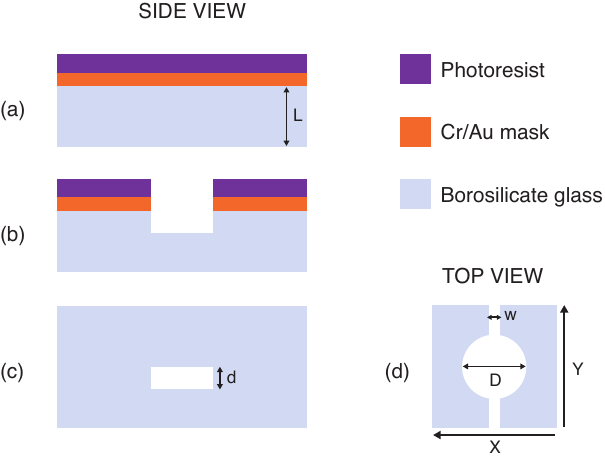}
\caption{Fabrication process of microfluidic cavities. (a) Deposition of a Cr/Au masking layer and photoresist on a piranha cleaned blank borosilicate glass substrate. (b) The design is transferred by optical lithography. The masking layer and the glass substrate are then etched. (c) Stripping of the photoresist and masking layer is followed by bonding with another blank borosilicate glass wafer. (d) Top view showing the resulting design.}
\label{nanofab_process}
\end{figure}

As described in Fig.\thinspace\ref{nanofab_process}, we start with borosilicate glass (Borofloat$^{\textregistered}$~33) wafers ($L=1.1$ mm) and, using standard photolithography, we transfer the design -- a circular basin ($D=5$ or 7 mm) with two rectangular channels ($w=1$ or 2 mm) leading from the edge of the device into the basin -- to the glass wafer. Then we successively wet etch the masking layers and the glass substrate down to a certain depth. This previous step defines the cavity depth which sets the confinement length for the liquid \he4. We present in this work the results obtained with two depths, $d\simeq$ 5 and 10 $\mu$m. It is interesting to note that the typical volume of liquid \he4 being studied in our microfluidic cavities is small ($D=7$ mm, $d=10\ \mu$m, $V \sim 1\ \mu$L) compared to those used in previous work~\cite{Barmatz1968, Rudnick1965, Lambert1979} on the first-sound velocity of liquid \he4. The wafer is then diced and bonded to a blank glass ``lid'' using direct bonding. This process involves cleaning the two pieces, first with a piranha solution (H$_2$SO$_4$ and H$_2$O$_2$), then with a soap solution and mechanical action. The two pieces are then pressed together by hand under a microscope. Sometimes, when the two surfaces are not perfectly clean the bonding does not work and one can see Newton's rings in the cavity due to light interference. In this case, the two pieces can easily be separated and the process repeated. Next, the samples are annealed at a temperature of $600^{\circ}\rm C$ for 2 hours to strengthen the bond. Finally, we affix (using Stycast$^{\textregistered}$ 1266) two broadband piezoelectric transducers of PZT-5A (Boston Piezo-Optics), one on each side of the microfluidic device. These transducers (diameter 4 mm, thickness 90 $\mu$m, with a fundamental compressional frequency of 20 MHz) have tab coaxial electrodes that allow both electrodes to be accessible from one side of the transducer. 

\begin{figure}[t]
\centering
\includegraphics[width=5cm]{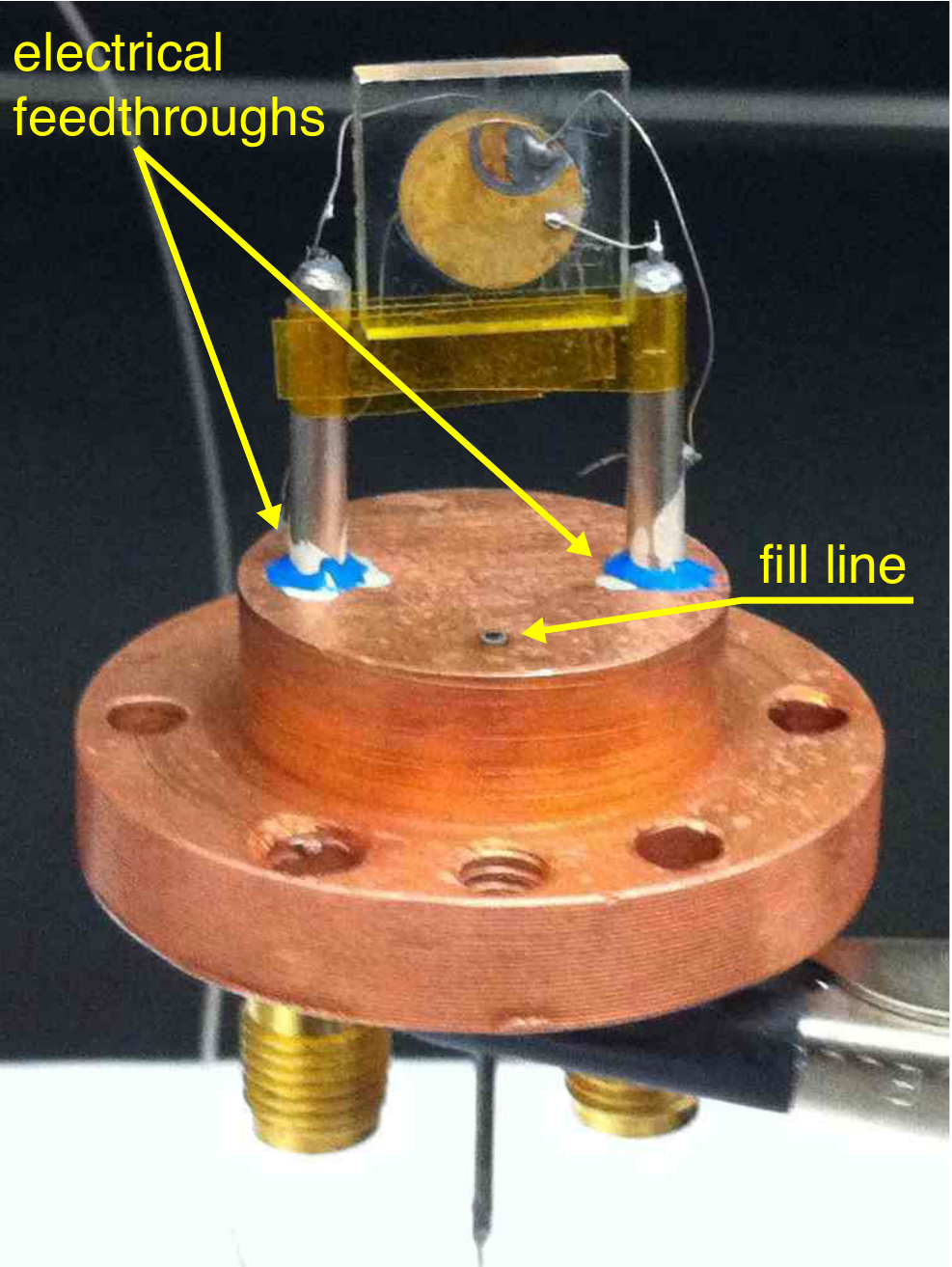}
\caption{Photograph of a sample mounted on the cap of the sample cell.}
\label{zoom_cap2}
\end{figure}

The electrodes of the transducers are connected to two coaxial feedthroughs placed on the cap of the sample cell (see Fig.\thinspace\ref{zoom_cap2}).  These feedthroughs and electrical connections are shielded to reduce capacitive cross-talk between the two drive electrodes ($< 0.5$ fF), allowing for separation of the ultrasonic signal from the capacitive background. For the coaxial cables connecting the feedthroughs to the top of the cryostat we used 50 $\Omega$ semi-rigid coaxial cable (model SC-086/50-SCN-CN from Coax Co.), which have low attenuation ($\sim1$ dB/m at 500 MHz) and thermal conductivity ($\sim7\times10^{-5}$ W$\cdot$cm/K) at 4 K. 

\begin{figure}[b]
\centering
\includegraphics[width=8cm]{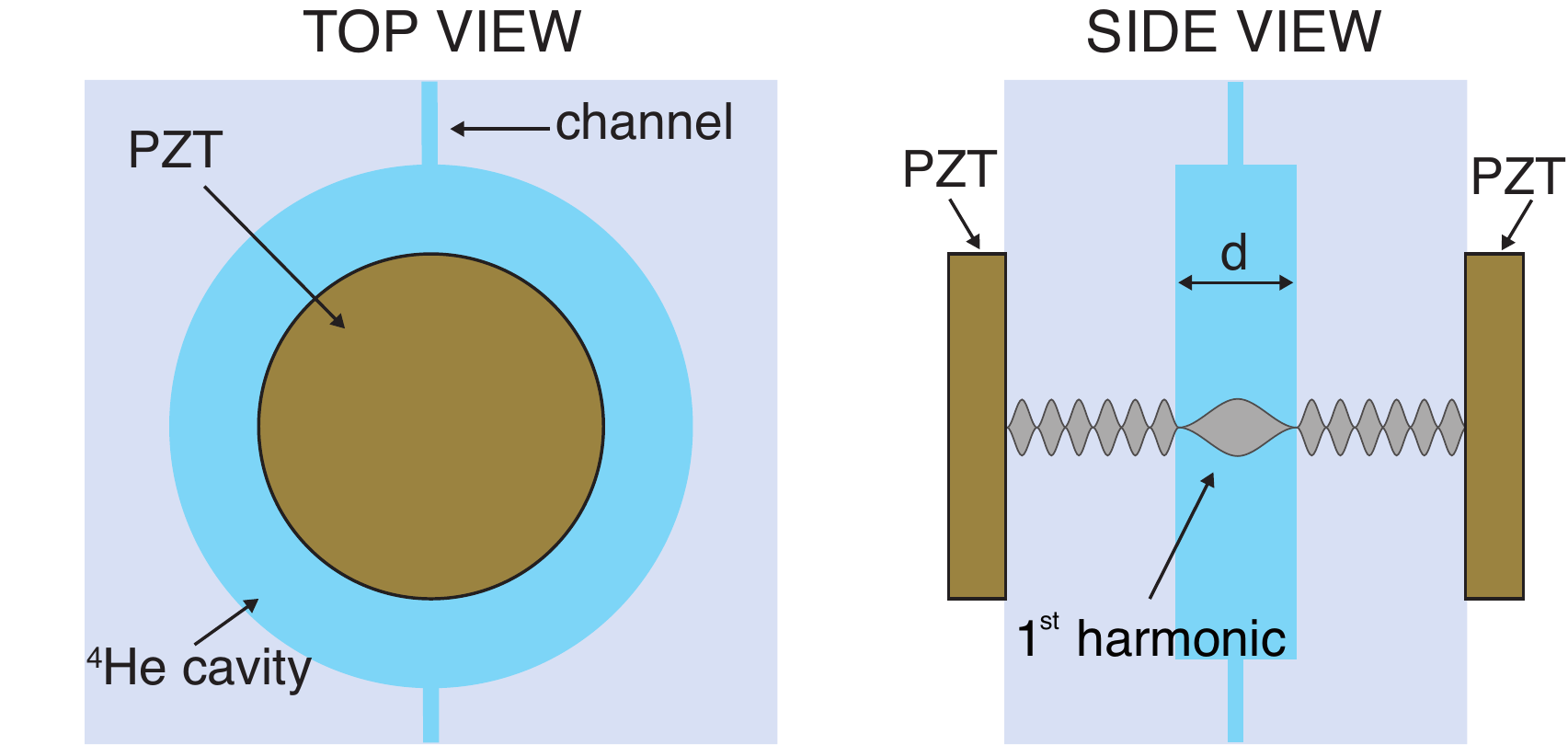}
\caption{Schematic of a microfluidic device with two PZT transducers bonded to it. The presence of the channels allows the microfluidic cavity to be filled with liquid \he4 (blue). The side view shows an ultrasonic standing wave (first harmonic) in the microfluidic cavity.}
\label{microfluidic_drawing_1st}
\end{figure}

By applying a drive voltage at MHz frequencies across the electrodes of the first transducer, ultrasonic waves are emitted that propagate across the sample,  reflect, and form standing waves in the microfluidic cavity.  As shown in Fig.~\ref{microfluidic_drawing_1st}, the transmission can be detected by a second transducer on the opposite side of the device.  The generation and detection of the voltage signals were performed using a 50 MHz lock-in amplifier (model HF2LI from Zurich Instruments) .

The completed UFPI is sealed into an oxygen free high conductivity copper (OFHC) sample cell with an indium O-ring, and immersed in liquid \he4.  This has the advantage that hydrostatic pressure is applied to the microfluidic sample, therefore avoiding any deformation of the cavity walls during pressurization (up to $\sim24.8$ bar). Figure~\ref{zoom_cap2} shows a microfluidic sample mounted on the sample cell, which contains two electrical feedthroughs and a fill line. The sample cell is attached to the 1 K pot of a commercial \he4 refrigerator via a teflon spacer and thermally anchored to the pot by a thin band of copper, resulting in low heat conduction between the two stages. This limits the cooling power demand on the 1 K pot when the sample cell is regulated at higher temperatures. A resistive carbon glass thermometer (model C12126 from Lakeshore) and a 50 $\Omega$ heater are also attached to the sample cell. The temperature is controlled with a precision of 50 $\mu$K by a proportional-integral-derivative (PID) controller on a resistance bridge (model 370 AC from LakeShore). Liquid pressure in the sample cell cavity is controlled with a precision of $\sim0.1$ \% using a gas handling system attached to a buffer cylinder. This experimental setup allows us to measure the first-sound of liquid \he4 in confined geometries as small as 5 $\mu$m, with fine resolution in both temperature and pressure.

\subsection{\label{sec:level2}Confinement characterization}
We used two different techniques to measure the cavity depth. The first was a surface profiler (Alpha Step IQ) prior to the bonding of the microfluidic samples. A diamond stylus, in contact with the sample, moved laterally across the sample to measure the topography. This tool gives the step height at a particular edge of the cavity with excellent resolution ($\sim$ 1 nm). However, it cannot be used to measure the depth profile over the entire cavity. In addition, this measurement cannot account for potential modifications of the depth induced by the bonding and annealing processes. 

The second technique used is based on the measurement of light interference in the microfluidic cavity. The devices are made out of glass and are transparent to visible light. When the microfluidic device is illuminated, the refractive index difference between glass ($n_{g}\simeq1.47$) and air ($n_{a}\simeq1$) causes the optical waves to be partially reflected at the glass-air interfaces. The reflected light interferes with the incident waves leading to interference patterns in the transmission. Thus the cavity depth profile can be obtained from measurement of the optical transmission through the cavity.

\begin{figure}[t]
\centering
\includegraphics[width=8cm]{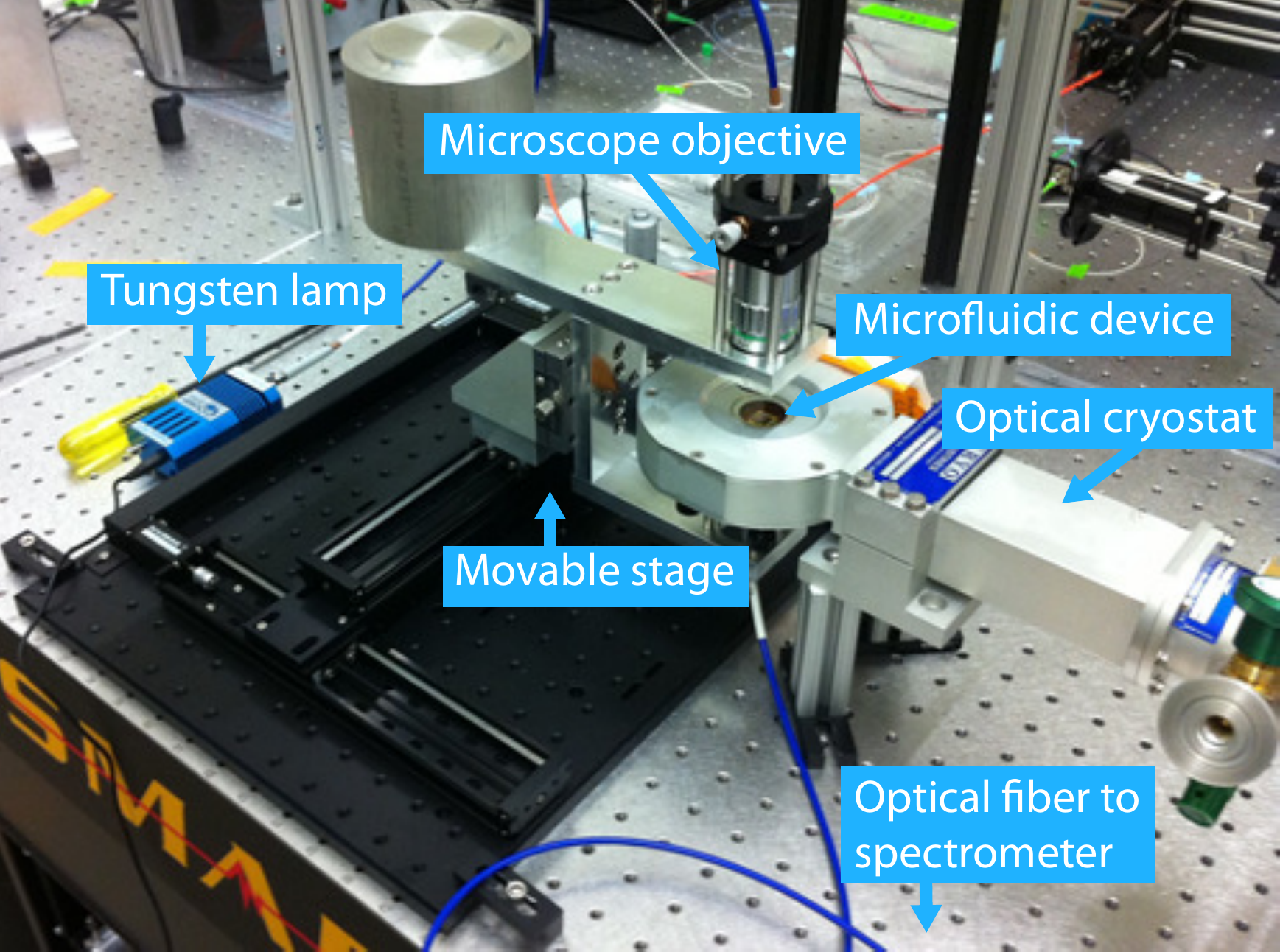}
\caption{Photograph of the experimental setup for optical interference mapping of the microfluidic cavity depth.}
\label{OpticalCryostat}
\end{figure}

We have used a broadband light source (LS1 tungsten halogen lamp from Ocean Optics), such that the coherence length of the light, $l_c$, is small compared to the glass thickness ($l_c \ll L$) but large enough that interference can still occur in the microfluidic cavity ($l_c > 2d$). This allows interference to occur solely in the microfluidic cavity, and not in the glass. Light from the lamp was sent into an optical fiber (model P600-1-SR from Ocean Optics) to illuminate the cavity through the objective of a microscope (model M Plan-Apo $20\times$ from Mitutoyo) for a spot size of $\sim30\ \mu$m in diameter. A second optical fiber was used to collect the light transmitted, guiding it to a compact CCD spectrometer (model CS200/M from Thorlabs) with a spectral resolution better than 2 nm (see Fig.~\ref{OpticalCryostat}). 

\begin{figure}[t]
\centering
\includegraphics[width=8cm]{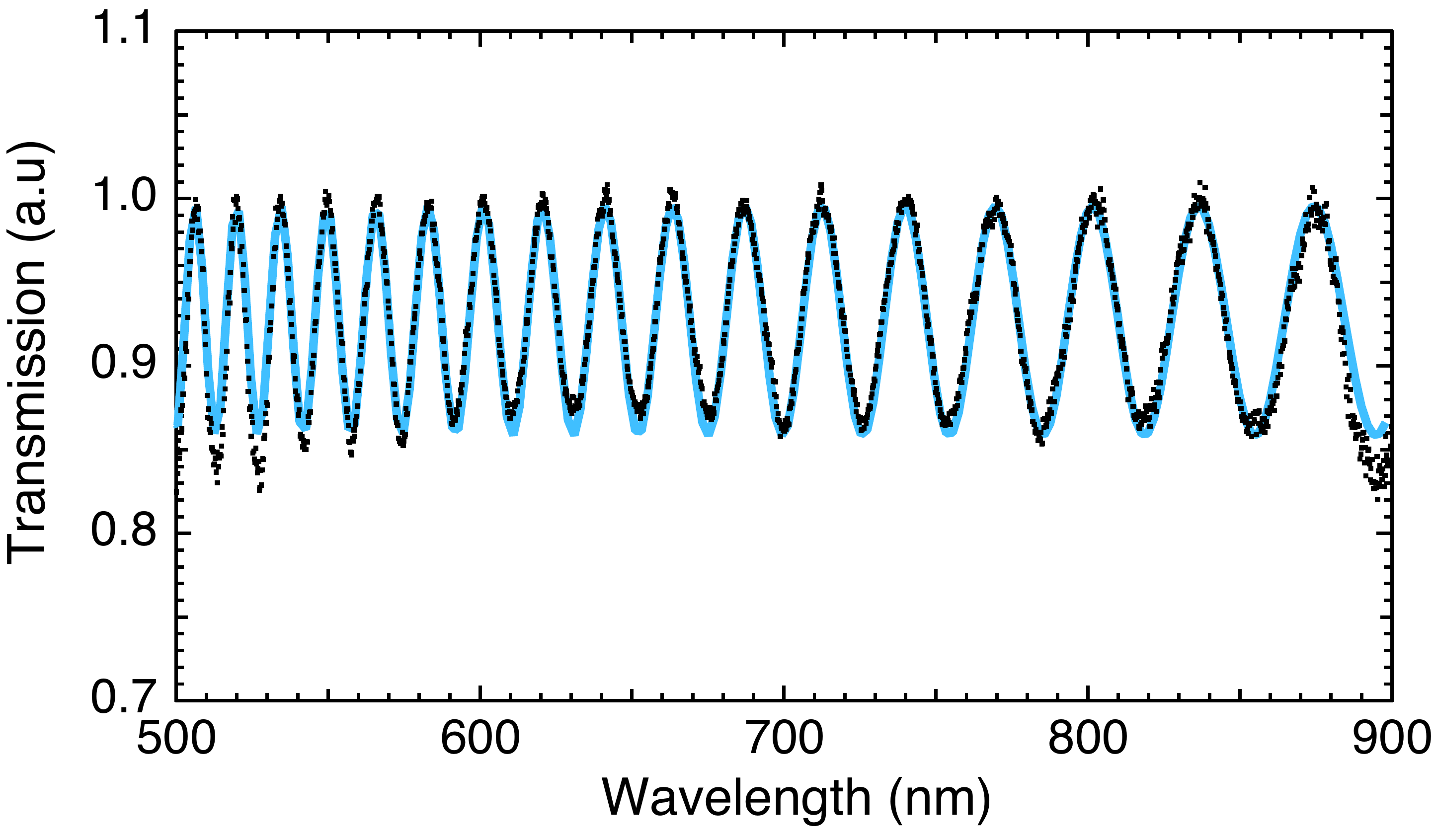}
\caption{Wavelength dependent transmission at one location of a microfluidic cavity (black dots). The fit to the spectral data (blue line) is obtained using Eq.~(\ref{eq:opticalFP}), and the best fit for $d$ at this location is $9.62\ \mu$m.}
\label{fit_optics_10um}
\end{figure}

By sampling the wavelength-dependent intensity of light transmitted through the cavity with the spectrometer, we can extract the size of the cavity from optical interference. In practice, we measure the transmission through two different paths. One path goes through the microfluidic cavity while the other one passes solely through the glass and is used as a reference. Hence, we separate the contribution of the microfluidic cavity from the glass substrate background. The small refractive index difference results in an optical cavity with low finesse $\mathcal{F}_o$,
\beq
\mathcal{F}_o=\frac{\pi}{2\times\arcsin{\left(\frac{1}{\sqrt{F_o}}\right)}}\sim0.17
\eeq
with the finesse coefficient $F_o=4R_o/(1-R_o)^2$ and the optical power reflection coefficient $R_o=\left|\frac{n_a-n_g}{n_a+n_g}\right|^2$. For this reason, along with the small coherence length of the light, it is a good approximation to only consider two beam interference in the microfluidic cavity. A two beam interference model in the cavity leads to the following expression for the optical transmission:
\beq\label{eq:opticalFP}
T_o = A+B*\cos\left({\frac{4\pi n_a d}{\lambda_o}}+\phi \right)
\eeq
with $\lambda_o$ the optical wavelength and $\phi$ an arbitrary phase. Ideally, if we know $n_a$ and $n_g$ precisely as the wavelength is varied we can fix both $A$ and $B$. However, in a real experiment, many factors arise (\emph{i.e.}~slightly misaligned optics, scattering, beam divergence, etc.) which will affect the total optical transmission. Therefore, in our experiment, we leave both $A$ and $B$ as free parameters to allow for a better fit of the depth $d$.  Fig.~\ref{fit_optics_10um} shows an example of the spectral data fit with Eq.~\ref{eq:opticalFP} at a particular location in a microfluidic cavity.

As discussed above, it is important to know the uniformity of the confinement geometry for the study of finite-size effects. We have been able to map the depth of our cavities by mounting the light source and the spectrometer on a two dimensional positioning  stage (model T-G-LSM200A200A from Zaber). In addition, to determine the temperature dependence of the cavity depth of our microfluidic cavities, we mounted them in an optical cryostat (model RC102-CFM from CryoIndustries).

\begin{figure}[t]
\centering
\includegraphics[width=8cm]{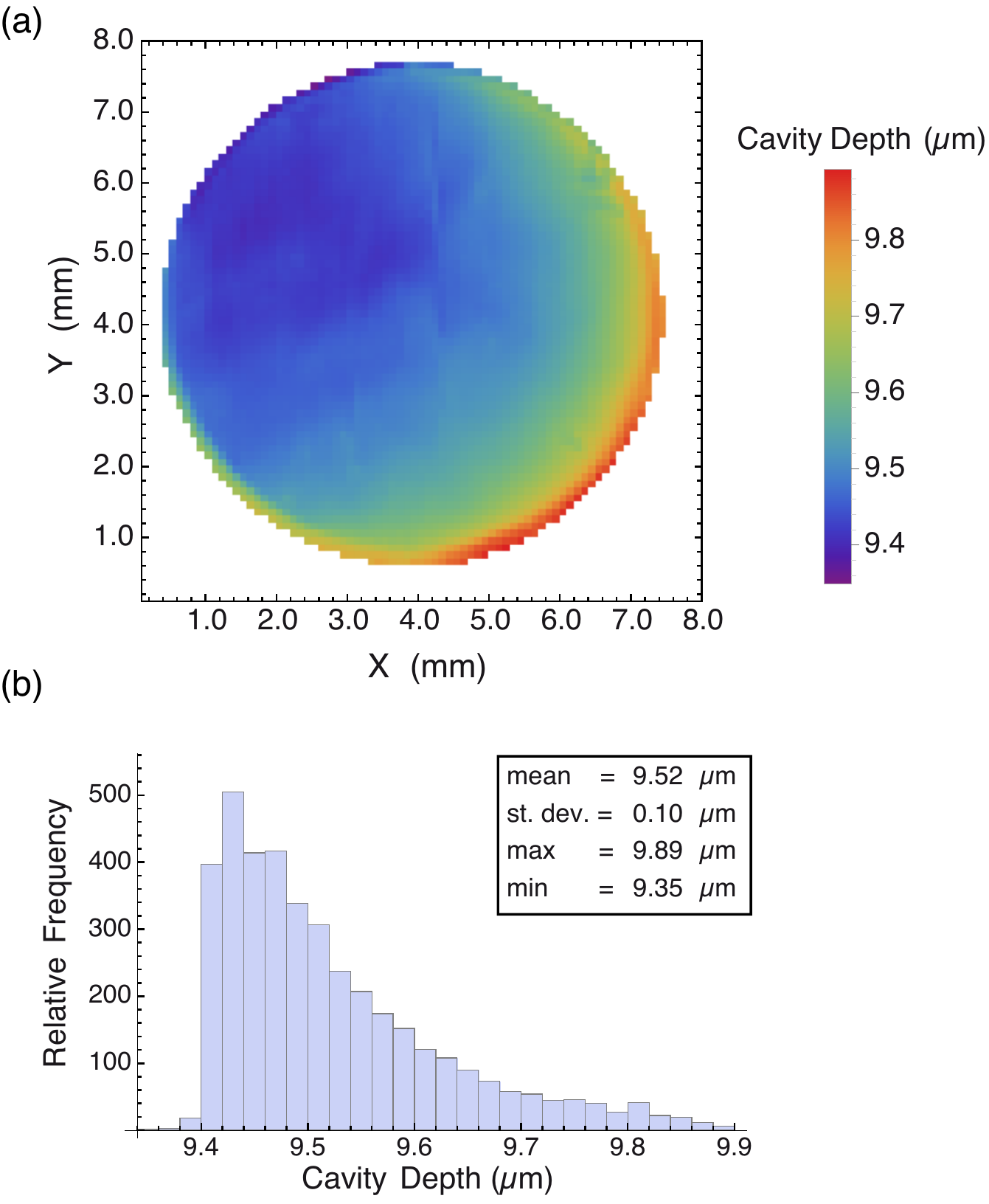}
\caption{(a) Map of a cavity depth profile measured in an optical cryostat at 77 K with a step size of 100 $\mu$m. (b) Histogram of the cavity depth profile showing the distribution of depth over the entire area of the cavity.}
\label{depthprofile&hist}
\end{figure}

We show in Fig.~\ref{depthprofile&hist} the cavity depth profile of a microfluidic cavity ($D=7$ mm, $d\simeq10\ \mu$m) at 77 K obtained with 100 $\mu$m steps in the $X$ and $Y$ directions. For this particular cavity, the average depth over the entire profile is 9.52 $\mu$m, which is in good agreement with the average value of the surface profiler (9.58 $\mu$m) measurements.  Moreover, using optical interferometry mapping allows us to quantify the depth uniformity over the cavity area. The spread in the depth ($\sigma=0.1\ \mu$m) represents only $\sim 1\%$ of the average cavity depth. For smaller cavities ($d_{\rm avg}=5.2\thinspace\mu$m), we obtained a similar spread of $\sim1.5$\% in the depth profile ($\sigma=0.082\thinspace\mu$m). We have noticed that this depth profile varies slightly from sample to sample and that the values furthest from the average are located near the edge of the cavity. We suspect that the depth profile of these cavities results from the non-uniformity of the wet etching in the fabrication process.

As for the temperature dependence of the cavity depth profile, measurements were performed at 300 and 77 K, and we found that any thermal contraction was smaller than our depth resolution and negligible compared to nonuniformities introduced during fabrication. This is in agreement with the very small values of the thermal expansion coefficients of most glass materials~\cite{Pobell2006}. Nevertheless, it was important to verify for our particular material (borosilicate glass) and geometry. As no significant thermal contraction is expected below 77 K, we did not explore this regime.

\subsection{\label{sec:level2_acoustifFP}Ultrasonic Fabry-P\'erot Interferometer}

The leading techniques for exploring acoustic properties of quantum liquids (\he3 and \he4) include measurements of the acoustic impedance \cite{Aoki2005,Davis2008,Murakawa2012} and propagation of ultrasonic pulses~\cite{Moon2010}. Neither of these are well suited to our measurement setup, because of the disruptive effects of the glass microfluidic cavity.  Instead, we have developed a new technique that benefits from the acoustic impedance mismatch by means of using the helium - glass interfaces as highly reflective acoustic mirrors of the resonant cavity of an ultrasonic Fabry-P\'erot interferometer.

The Fabry-P\'erot interferometer is based on wave interference in a resonant cavity with reflective walls. The acoustic transmission through the Fabry-P\'erot interferometer for a normally incident acoustic beam, is given by
\beq \label{eq:acousticFP}
T_a=\frac{1}{1+\mathcal{F}_a\sin^2{(\delta\phi/2)}}
\eeq
with $\delta\phi=4\pi d / \lambda_a $ the phase difference between two successive transmitted waves and $\lambda_a$ the acoustic wavelength. For large reflectivity ($R_a>0.5$), the finesse $\mathcal{F}_a$ can be approximated as 
\beq
\mathcal{F}_a\simeq\frac{\pi R_a^{1/2}}{1-R_a},
\eeq
with $R_a$ the acoustic reflection coefficient of the walls of the cavity. Considering only the propagation of longitudinal acoustic waves, the walls of the cavity are the liquid \he4 - glass interfaces and the reflection coefficient is
\beq
R_a=\left|\frac{Z_0^g-Z_0^h}{Z_0^g+Z_0^h}\right|^2.
\eeq
Here $Z_0^g=\rho_g v_g$ ($Z_0^h=\rho_h v_h$) is the characteristic acoustic impedance of borosilicate glass (liquid \he4), $\rho_j$ is the density and $v_j$ is the longitudinal velocity of the materials. It turns out that the acoustic impedance of liquid \he4 ($Z_0^h=3.3\times10^4$ kg m$^{-2}$s$^{-1}$ at 2 K and saturated vapor pressure) is smaller by more than two orders of magnitude than that of the borosilicate glass ($Z_0^g=1.3\times10^7$ kg m$^{-2}$s$^{-1}$). This leads to a large reflection coefficient ($R_a=99$\%) for the liquid \he4 - glass interfaces, which then act as highly reflective acoustic mirrors. As a result, the expected finesse of the acoustic cavity should be relatively high ($\mathcal{F}_a\simeq284$). Following Eq.~\ref{eq:acousticFP}, the transmission will show peaks when $\lambda_a=2d/n$, with $n$ a positive integer. Hence, the resonance frequencies of the acoustic cavities are given by
\beq\label{eq:freq_vel}
f_n=n\frac{v_h}{2d}.
\eeq
A calculation for a microfluidic cavity of thickness $d=9.52\ \mu$m filled with liquid \he4 of sound velocity $v_h=227$ m/s at 2 K leads to a free spectral range $\Delta f= f_{n+1}-f_{n}\simeq11.92$~MHz.  By measuring the transmission through the microfluidic device, one should be able to measure the resonance peak and, knowing the cavity depth, should be able to extract the first-sound velocity of the confined liquid \he4.

\begin{figure}[t]
\centering
\includegraphics[width=8cm]{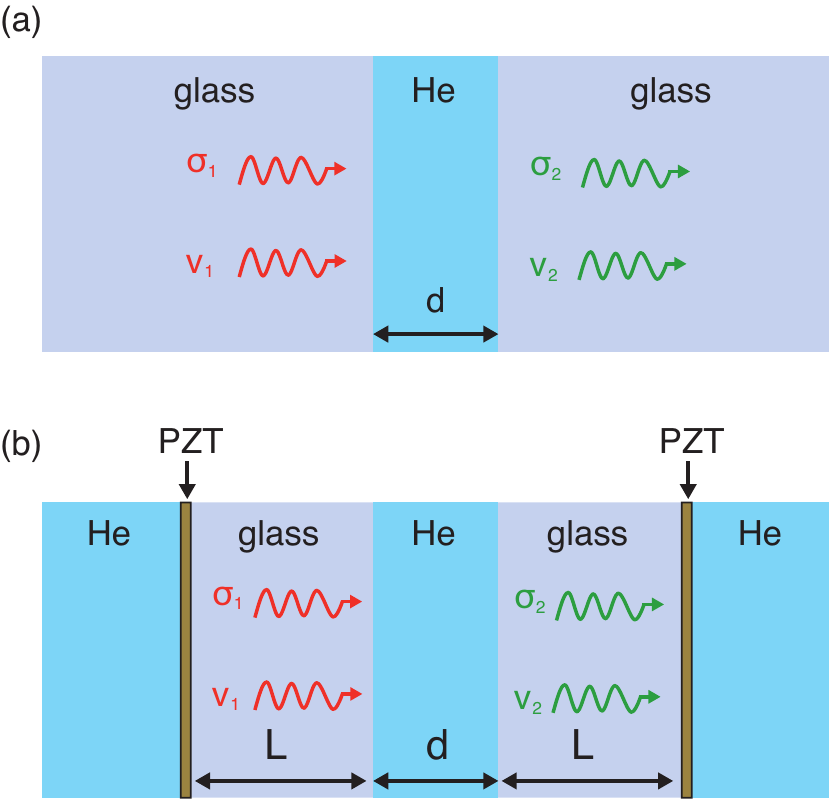}
\caption{(a) Propagation of acoustic waves through a single slab of helium surrounded by an infinite medium of glass. (b) Propagation through three consecutive slabs, glass, helium and glass, bounded by two PZT.}
\label{MatrixTransfer_schematic}
\end{figure}

This model helps to understand the principle of our ultrasonic Fabry-P\'erot interferometer, however, to describe our data accurately we need to use a more complete theory. Above, we have treated the cavity as being surrounded by an infinite medium made of glass. In reality, as illustrated in Fig.~\ref{MatrixTransfer_schematic}, the glass wafers, on each side of the cavity, have a finite thickness ($L=1.1$ mm). In addition, there are piezoelectric transducers (PZT) bonded to each side of the device, which is immersed in liquid \he4. In order to account for the finite size of these different layers, we have used a transfer matrix formalism originating from the 1D Mason model~\cite{Cai2001}.

In this formalism, the transmission of acoustic waves through a layer of finite dimension is expressed in terms of a stress ($\sigma_j$) and velocity ($v_j$). The stress-velocity vector ($\sigma_2,v_2$) after each layer is expressed as a function of the vector before that layer ($\sigma_1,v_1$) via a transfer matrix
\beq
\left(
\begin{array}{c}
\sigma_2 \\
v_2 \\
\end{array}
\right) =
M_j\cdot\left(
\begin{array}{c}
{\sigma_1} \\
{v_1} \\
\end{array}
\right).
\eeq
For a given material of characteristic acoustic impedance $Z_0^j$, thickness $d_j$ and acoustic wave vector $k_j=2\pi/\lambda_j$, the transfer matrix is :
\beq
M_j=
\left(
\begin{array}{cc}
\cos{(k_jd_j)} &\mathrm{i}\sin{(k_jd_j)}  \\
\frac{\mathrm{i}}{ Z_{0}^j}\sin{(k_jd_j)} & \cos{(k_jd_j)}  \\
\end{array}
\right),
\eeq
This formalism not only allows one to find the acoustic transmission for a single layer (Eq.~\ref{eq:acousticFP}) but offers the possibility to  compute the transmission through a series of layers by multiplying the transfer matrices of each layer together. 

In our study, we neglect the PZT layers, which contribute only in enhancing the ultrasonic signal near their fundamental resonance ($\sim20$ MHz). We only consider the system of the three layers in series: glass - liquid \he4 - glass. The total transfer matrix is obtained by multiplying the transfer matrix of each layer together,
\beq
M_{\rm tot}=M_g\cdot M_h\cdot M_g,
\eeq
with $M_g$ ($M_h$) the transfer matrix of the glass (liquid \he4) layers. Using the linear relation, 
\beq
\left(
\begin{array}{c}
\sigma_2 \\
v_2 \\
\end{array}
\right)=
M_{\rm tot}
\left(
\begin{array}{c}
{\sigma_1} \\
{v_1} \\
\end{array}
\right),
\eeq
we calculate the total acoustic impedance $Z_{\rm tot}$ of the system of the three layers in series to be
\beq
Z_{\rm tot}=-\frac{\sigma_2}{v_2}.
\eeq
The acoustic transmission through the system of the three layers in series, immersed in liquid \he4 is
\beq
T_a=1-\left|\frac{Z_{\rm tot}-Z_{0}^h}{Z_{\rm tot}+Z_{0}^h}\right|^2.
\eeq

\begin{figure}[t]
\centering
\includegraphics[width=8cm]{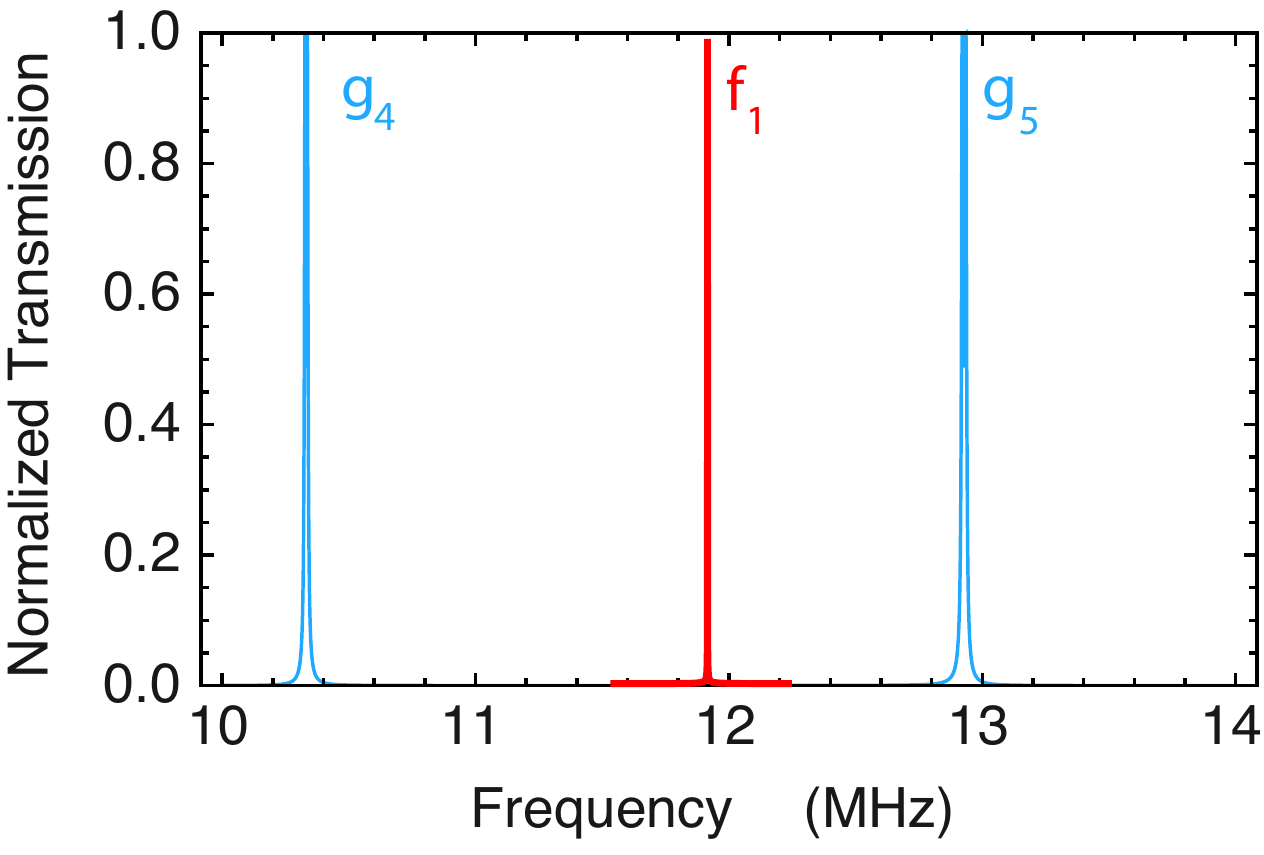}
\caption{Calculation of the acoustic transmission through a microfluidic device ($L=1.1$ mm and $d=9.52\ \mu$m) using the transfer matrix formalism. $f_1$ (red line) is the first harmonic of the resonant cavity filled with liquid \he4 with a sound velocity of 227 m/s (value at 2 K). The $g_n$ (blue lines) are the $n^{\rm th}$ harmonics in the glass layers with a sound velocity of 5685 m/s.}
\label{TransferMat_graph}
\end{figure}

We show in Fig.\thinspace\ref{TransferMat_graph} the computed acoustic transmission through a typical microfluidic device ($L=1.1$ mm and $d=9.52\ \mu$m) as a function of the ultrasonic frequency. This graph shows two resonance peaks in the glass ($g_n$), and the resonance peak corresponding to the first harmonic ($f_1$) of the microfluidic cavity filled with liquid \he4. In this case, the resonance frequencies in the glass are approximately given by
\beq
g_n\simeq n\frac{v_g}{2L},
\eeq
leading to a free spectral range of $\Delta g\simeq2.58$ MHz with $v_g=5685$ m/s the sound velocity in borosilicate glass. When the cavity resonance peaks are sufficiently far from the glass resonances, it is a good approximation to use Eq.~\ref{eq:freq_vel} to calculate the sound velocity of liquid \he4.
\begin{figure}[t]
\centering
\includegraphics[width=7.5cm]{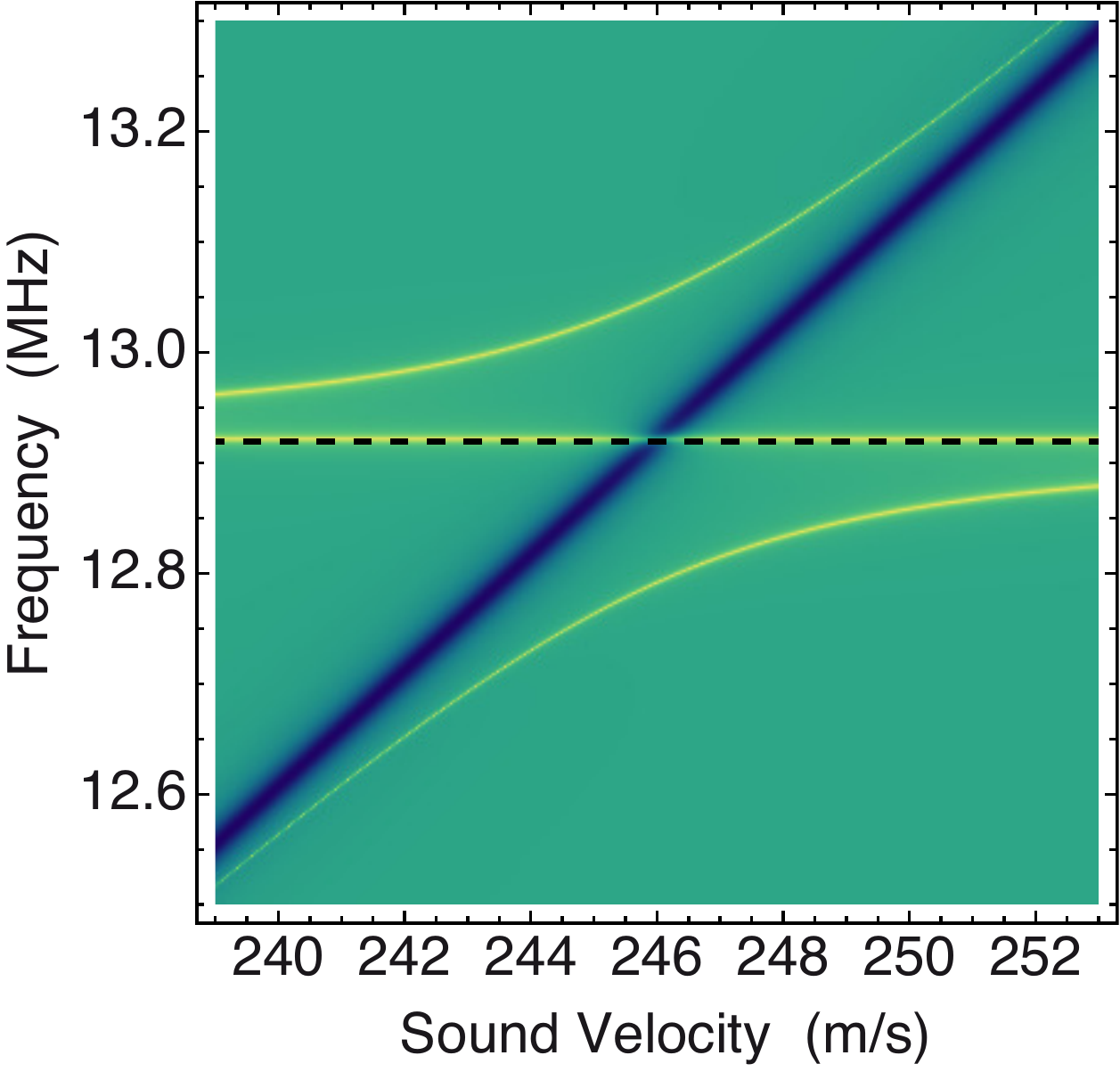}
\caption{Calculation of the transmission through a microfluidic device ($L=1.1$ mm and $d=9.52\ \mu$m) as a function of the frequency and sound velocity in liquid \he4. A pure cavity mode ($n=1$) is obtained using Eq.\thinspace(\ref{eq:freq_vel}) (thick blue line). The hybridization of this mode (yellow line) with a glass mode (dashed line) is obtained using the matrix transfer formalism for layers in series (glass - liquid \he4 - glass).}
\label{Hybridized_modes}
\end{figure}
On the other hand, when a cavity resonance ($f_n$) approaches a glass resonance ($g_n$), the two modes start to hybridize, as illustrated in Fig.~\ref{Hybridized_modes}. We show in Fig.~\ref{Hybridized_modes} a calculation of the transmission through a microfluidic device ($L=1.1$ mm and $d=9.52\ \mu$m) zoomed around $g_5\simeq12.9$ MHz. Eq.\thinspace\ref{eq:freq_vel} leads to a single line as a function of the sound velocity in liquid \he4, while a calculation taking into account the finite size of the glass shows hybridization of the two modes. Due to this hybridization, it is more difficult to extract the sound velocity from the frequency spectrum of the transmission. Experimentally, we have been able to decrease the effect of the hybridization by lowering the finesse of the glass resonances, through scratching the glass surfaces. As a result, it is possible to track the \he4 cavity resonance peaks as a function of temperature and pressure, and extract the sound velocity of the confined liquid \he4.

\section{\label{sec:level1}Results and Discussion}
We measure the frequency spectrum of the transmission through different microfluidic devices by using the following method. A drive voltage ($V_d\sim100$ mV) swept in frequency (10 to 50 MHz), is applied to one piezoelectric transducer. The ultrasonic signal after traversing the microfluidic sample is received by the second transducer. The response voltage is measured ($V_m\sim$ 0.01 - 15 mV) by the lock-in amplifier. In our first measurements, the transmission spectrum showed the resonance peaks of the glass layers with higher amplitude than the cavity resonance peaks. Experimentally, the glass layers act as parasitic filters for the ultrasonic transmission and the cavity resonance peaks are difficult to observe unless they are close enough to glass resonance peaks. In this configuration, it is hard to track the cavity resonance peaks. 

\begin{figure}[t]
\centering
\includegraphics[width=8cm]{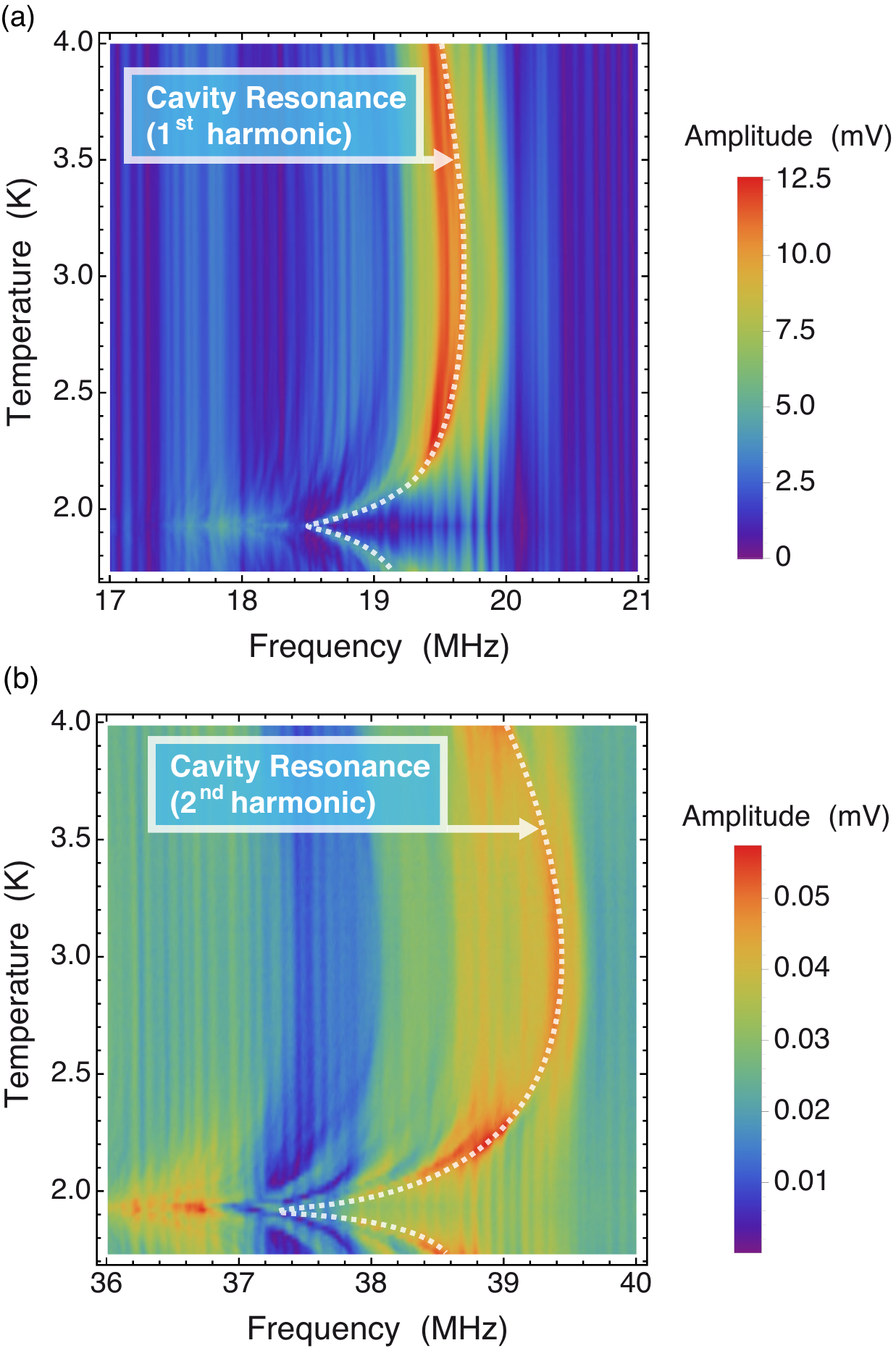}
\caption{Cavity resonance peaks (white dashed lines) of a microfluidic device ($L=1.1$ mm and $d=9.52\thinspace\mu$m) filled with liquid \he4 at a constant pressure $P=20.80\pm0.02$ bar. (a) Cavity first harmonic ($f_1$). (b) Cavity second harmonic ($f_2$).}
\label{Device8_mode1&2_20.8bar}
\end{figure}
In order to improve both the contrast in signal amplitude and the tracking of the cavity resonance peaks, we reduced the finesse of the glass modes by roughening the surface of the microfluidic device prior to the bonding of the transducer. By scratching the surface of the devices with varying grits of sandpaper we drastically reduced the finesse of the glass resonances and widen the bandwidth of these parasitic filters. Even though we could not suppress the glass resonances completely, the contrast between the two peaks was greatly improved and we could track the temperature dependence of the cavity peaks for different microfluidic devices at different pressures.

Figure~\ref{Device8_mode1&2_20.8bar} shows the temperature dependence (from 4 to 1.7 K) of the cavity resonance frequencies ($f_1$ and $f_2$) for a cavity ($d=9.52$ $\mu$m) filled with liquid \he4 at 20.80 $\pm$ 0.02 bar. We can see from this graph that the measured amplitude ranges from 0 to 12.5 mV for the first harmonic ($f_1\sim19$ MHz) while it is smaller (0 to 0.06 mV) for the second harmonic ($f_2\sim39$ MHz). This is due to the finite bandwidth of our piezoelectric transducers, which have a higher sensitivity around their resonance frequency ($\sim20$ MHz). In our experiment, the sensitivity of the transducer was a significant limiting factor for the study of higher harmonics and smaller cavity depths.

To extract the sound velocity from our measurement, while accounting for the hybridization between the glass and cavity modes, we use an effective cavity depth, $d_{\rm eff}$. We find that a single effective cavity depth ($d_{\rm eff}=8.85 \pm 0.04\ \mu$m) accounts for all of our data, although this is slightly different from the measured depth of the microfluidic cavity ($d=9.52\thinspace\mu$m).  The exact mechanism that leads to this effective depth is difficult to pinpoint, since it is the result of a complex coupling interaction between the glass resonance background and the cavity resonance peaks. Nevertheless, it is noteworthy that using a single effective cavity depth accounts for both helium cavity harmonics, $f_1$ and $f_2$, and leads to consistent sound velocity temperature and pressure dependence.  This confirms that the effective depth depends only on the geometry of the microfluidic sample. 

To investigate this further, we found that by altering the thickness of the glass layers can we change this effective cavity depth.  Specifically, we have used two silica acoustic delay rods (length 10 mm and diameter 5 mm) bonded between the glass wafers and the transducers, resulting in an effective glass thickness of $L=11.1$ mm.  We performed measurements on these devices and extracted an effective cavity depth of $d_{\rm eff} = 11.85\thinspace\mu$m -- larger than the measured cavity depth, further confirming the effective cavity depth is only a property of the device geometry and does not reveal new physics of the liquid helium. 

\begin{figure}[t]
\centering
\includegraphics[width=8.5cm]{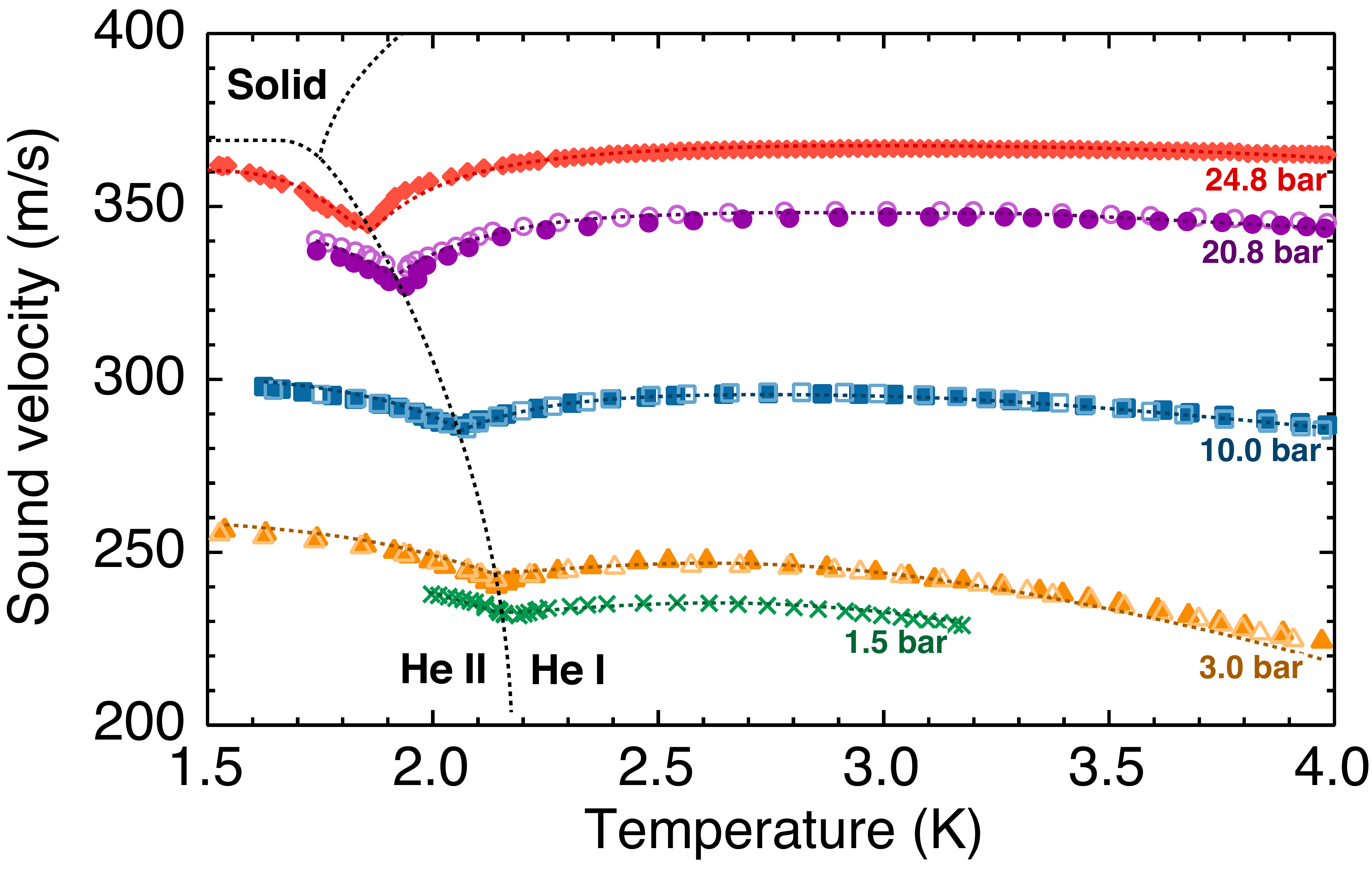}
\caption{Sound velocity temperature dependence (1.5 - 4.0 K) at different pressures (3.0 - 24.8 bar) of liquid \he4 confined in a microfluidic cavity ($d=9.52\thinspace\mu$m). Data extracted from the first (full symbols) and the second harmonic (empty symbols) of the resonant helium cavity of the ultrasonic Fabry-P\'erot interferometer. We also show data (cross symbols) for the measurement of the first harmonics in a $5.18\thinspace\mu$m cavity. Data are compared with bulk sound velocity (colored dashed lines) measured by Vignos \textit{et al.}~\cite{Vignos1966} and corrected with more accurate temperature values from Maynard \textit{et al.}~\cite{Maynard1976}. The phase boundaries between the different phases (dashed lines) are shown as a guide to the eye.}
\label{device8_allcurves}
\end{figure}

With the effective cavity depth and the cavity resonance frequencies ($f_1$ and $f_2$), we calculate the first-sound velocity $v_h$ of the layer of confined liquid \he4, using Eq.\thinspace\ref{eq:freq_vel}.  We show in Fig.~\ref{device8_allcurves} the experimental first-sound velocity at different pressures, measured for a microfluidic cavity of average depth $d=9.52\thinspace\mu$m.  We have also been able to measure the sound velocity of liquid \he4 confined in a cavity with a depth as small as $d=5.18\ \mu$m ($d_{\rm eff}=4.32\ \mu$m), shown as the crosses in Fig.\thinspace\ref{device8_allcurves}.

It is important to note that for a cavity of about 10 $\mu$m, finite-size effects should appear\cite{Gasparini2008} at a temperature $T$ such that $|T-T_\lambda|<1\ \mu$K. Thus, we expect our sound velocity data to be consistent with the bulk values for the entire temperature range (1.5 - 4.0 K), except extremely close to $T_\lambda$. Our resolution in temperature (50 $\mu$K) does not allow us to currently observe these effects. We compare our results with the experimental bulk liquid \he4 data measured by Vignos \textit{et al.}~\cite{Vignos1966}, which we have corrected with more precise temperature values measured by Maynard \textit{et al.}~\cite{Maynard1976}. As seen in Fig.\thinspace\ref{device8_allcurves}, the agreement in the temperature and pressure dependence between our data and the bulk value is very good. We obtained this agreement using an effective depth for the cavities. Further theoretical studies on the acoustic interaction between the glass layers and the cavity could help to predict if the finite-size effects in thinner cavities will be governed by the geometric depth or the effective depth. In the future, it will be possible to first characterize the effective depth of thinner cavities with respect to the known bulk behavior, and then extract the finite-size effect contribution appearing very near $T_{\lambda}$. Moreover, we will be able to identify which one of these depths is responsible for finite-size effects, by changing the thickness of the glass layer while keeping the depth of the cavity constant.

Our apparatus is ideal for the study of critical ultrasonic behavior~\cite{Bhattacharyya1998} in finite geometries as it allows  study of different frequencies via the harmonics of the cavity. However, better resolution in the ultrasonic attenuation is required, which could be achieved by further modifying the substrate of the microfluidic sample.

\section{\label{sec:level1}Conclusion}

We have designed and tested an ultrasonic Fabry-P\'erot interferometer in order to probe finite-size effects in the first-sound of liquid \he4 near the superfluid transition.  We have been able to measure for the first time the first-sound of liquid \he4 in cavities as small as 9.52 and 5.18 $\mu$m. These devices were fully characterized using profilometer and scanning optical interference techniques.  An important advantage of our experimental setup is that our devices are insensitive to liquid \he4 outside the microfluidic cavity and therefore we are able to work at hydrostatic pressure.  As a result we have demonstrated working pressures as high as 24.8 bar, much greater than previous systems, without causing deformation of the confinement geometry.

The results on the first-sound velocity obtained in our microfluidic cavities agree well with the sound velocities previously measured in the bulk limit.  Future work on smaller cavities, with higher frequencies piezoelectric transducers, are anticipated to reveal finite-size effects in liquid \he4, and provide a complimentary technique to the extensive work on heat capacity in this system.  

\section{\label{sec:level1}Acknowledgements}
This work was supported by the University of Alberta, Faculty of Science; the Natural Sciences and Engineering Research Council, Canada; the Canada Foundation for Innovation; and Alberta Innovates Technology Futures. We would like to thank J. R. Beamish and S. Balibar for stimulating discussions, and F. Hegmann for assistance with optical characterization of microcavities. We are grateful to the technical support of G. Popowich and D. Mullin, and the staff of the University of Alberta NanoFab for their assistance in device fabrication.

\bibliography{apssamp}

\providecommand{\noopsort}[1]{}\providecommand{\singleletter}[1]{#1}%
\begin{thebibliography}{43}%
\makeatletter
\providecommand \@ifxundefined [1]{%
 \@ifx{#1\undefined}
}%
\providecommand \@ifnum [1]{%
 \ifnum #1\expandafter \@firstoftwo
 \else \expandafter \@secondoftwo
 \fi
}%
\providecommand \@ifx [1]{%
 \ifx #1\expandafter \@firstoftwo
 \else \expandafter \@secondoftwo
 \fi
}%
\providecommand \natexlab [1]{#1}%
\providecommand \enquote  [1]{``#1''}%
\providecommand \bibnamefont  [1]{#1}%
\providecommand \bibfnamefont [1]{#1}%
\providecommand \citenamefont [1]{#1}%
\providecommand \href@noop [0]{\@secondoftwo}%
\providecommand \href [0]{\begingroup \@sanitize@url \@href}%
\providecommand \@href[1]{\@@startlink{#1}\@@href}%
\providecommand \@@href[1]{\endgroup#1\@@endlink}%
\providecommand \@sanitize@url [0]{\catcode `\\12\catcode `\$12\catcode
  `\&12\catcode `\#12\catcode `\^12\catcode `\_12\catcode `\%12\relax}%
\providecommand \@@startlink[1]{}%
\providecommand \@@endlink[0]{}%
\providecommand \url  [0]{\begingroup\@sanitize@url \@url }%
\providecommand \@url [1]{\endgroup\@href {#1}{\urlprefix }}%
\providecommand \urlprefix  [0]{URL }%
\providecommand \Eprint [0]{\href }%
\providecommand \doibase [0]{http://dx.doi.org/}%
\providecommand \selectlanguage [0]{\@gobble}%
\providecommand \bibinfo  [0]{\@secondoftwo}%
\providecommand \bibfield  [0]{\@secondoftwo}%
\providecommand \translation [1]{[#1]}%
\providecommand \BibitemOpen [0]{}%
\providecommand \bibitemStop [0]{}%
\providecommand \bibitemNoStop [0]{.\EOS\space}%
\providecommand \EOS [0]{\spacefactor3000\relax}%
\providecommand \BibitemShut  [1]{\csname bibitem#1\endcsname}%
\let\auto@bib@innerbib\@empty
\bibitem [{\citenamefont {Duh}\ \emph {et~al.}(2012)\citenamefont {Duh},
  \citenamefont {Suhel}, \citenamefont {Hauer}, \citenamefont {Saeedi},
  \citenamefont {Kim}, \citenamefont {Biswas},\ and\ \citenamefont
  {Davis}}]{Duh2012}%
  \BibitemOpen
  \bibfield  {author} {\bibinfo {author} {\bibfnamefont {A.}~\bibnamefont
  {Duh}}, \bibinfo {author} {\bibfnamefont {A.}~\bibnamefont {Suhel}}, \bibinfo
  {author} {\bibfnamefont {B.}~\bibnamefont {Hauer}}, \bibinfo {author}
  {\bibfnamefont {R.}~\bibnamefont {Saeedi}}, \bibinfo {author} {\bibfnamefont
  {P.}~\bibnamefont {Kim}}, \bibinfo {author} {\bibfnamefont {T.}~\bibnamefont
  {Biswas}}, \ and\ \bibinfo {author} {\bibfnamefont {J.}~\bibnamefont
  {Davis}},\ }\href
  {http://link.springer.com/article/10.1007%2Fs10909-012-0617-4} {\bibfield
  {journal} {\bibinfo  {journal} {J. Low Temp. Phys.}\ }\textbf {\bibinfo
  {volume} {168}},\ \bibinfo {pages} {31} (\bibinfo {year} {2012})}\BibitemShut
  {NoStop}%
\bibitem [{\citenamefont {Levitin}\ \emph {et~al.}(2013)\citenamefont
  {Levitin}, \citenamefont {Bennett}, \citenamefont {Casey}, \citenamefont
  {Cowan}, \citenamefont {Saunders}, \citenamefont {Drung}, \citenamefont
  {Schurig},\ and\ \citenamefont {Parpia}}]{Levitin2013}%
  \BibitemOpen
  \bibfield  {author} {\bibinfo {author} {\bibfnamefont {L.~V.}\ \bibnamefont
  {Levitin}}, \bibinfo {author} {\bibfnamefont {R.~G.}\ \bibnamefont
  {Bennett}}, \bibinfo {author} {\bibfnamefont {A.}~\bibnamefont {Casey}},
  \bibinfo {author} {\bibnamefont {Cowan}}, \bibinfo {author} {\bibfnamefont
  {J.}~\bibnamefont {Saunders}}, \bibinfo {author} {\bibfnamefont
  {D.}~\bibnamefont {Drung}}, \bibinfo {author} {\bibfnamefont
  {T.}~\bibnamefont {Schurig}}, \ and\ \bibinfo {author} {\bibfnamefont
  {J.~M.}\ \bibnamefont {Parpia}},\ }\href
  {http://www.sciencemag.org/content/340/6134/841} {\bibfield  {journal}
  {\bibinfo  {journal} {Science}\ }\textbf {\bibinfo {volume} {340}},\ \bibinfo
  {pages} {841} (\bibinfo {year} {2013})}\BibitemShut {NoStop}%
\bibitem [{\citenamefont {Vorontsov}\ and\ \citenamefont
  {Sauls}(2007)}]{Vorontsov2007}%
  \BibitemOpen
  \bibfield  {author} {\bibinfo {author} {\bibfnamefont {A.~B.}\ \bibnamefont
  {Vorontsov}}\ and\ \bibinfo {author} {\bibfnamefont {J.~A.}\ \bibnamefont
  {Sauls}},\ }\href {http://link.aps.org/doi/10.1103/PhysRevLett.98.045301}
  {\bibfield  {journal} {\bibinfo  {journal} {Phys. Rev. Lett.}\ }\textbf
  {\bibinfo {volume} {98}},\ \bibinfo {pages} {045301} (\bibinfo {year}
  {2007})}\BibitemShut {NoStop}%
\bibitem [{\citenamefont {Sauls}(2011)}]{Sauls2011}%
  \BibitemOpen
  \bibfield  {author} {\bibinfo {author} {\bibfnamefont {J.~A.}\ \bibnamefont
  {Sauls}},\ }\href {http://link.aps.org/doi/10.1103/PhysRevB.84.214509}
  {\bibfield  {journal} {\bibinfo  {journal} {Phys. Rev. B}\ }\textbf {\bibinfo
  {volume} {84}},\ \bibinfo {pages} {214509} (\bibinfo {year}
  {2011})}\BibitemShut {NoStop}%
\bibitem [{\citenamefont {Wu}\ and\ \citenamefont {Sauls}(2013)}]{Wu2013}%
  \BibitemOpen
  \bibfield  {author} {\bibinfo {author} {\bibfnamefont {H.}~\bibnamefont
  {Wu}}\ and\ \bibinfo {author} {\bibfnamefont {J.~A.}\ \bibnamefont {Sauls}},\
  }\href {http://link.aps.org/doi/10.1103/PhysRevB.88.184506} {\bibfield
  {journal} {\bibinfo  {journal} {Phys. Rev. B}\ }\textbf {\bibinfo {volume}
  {88}},\ \bibinfo {pages} {184506} (\bibinfo {year} {2013})}\BibitemShut
  {NoStop}%
\bibitem [{\citenamefont {Wiman}\ and\ \citenamefont
  {Sauls}(2013)}]{Wiman2013}%
  \BibitemOpen
  \bibfield  {author} {\bibinfo {author} {\bibfnamefont {J.~J.}\ \bibnamefont
  {Wiman}}\ and\ \bibinfo {author} {\bibfnamefont {J.~A.}\ \bibnamefont
  {Sauls}},\ }\href
  {http://link.springer.com/article/10.1007%2Fs10909-013-0924-4} {\bibfield
  {journal} {\bibinfo  {journal} {J. Low Temp. Phys.}\ }\textbf {\bibinfo
  {volume} {175}},\ \bibinfo {pages} {17} (\bibinfo {year} {2013})}\BibitemShut
  {NoStop}%
\bibitem [{\citenamefont {Garcia}\ and\ \citenamefont
  {Chan}(1999)}]{Garcia1999}%
  \BibitemOpen
  \bibfield  {author} {\bibinfo {author} {\bibfnamefont {R.}~\bibnamefont
  {Garcia}}\ and\ \bibinfo {author} {\bibfnamefont {M.~H.~W.}\ \bibnamefont
  {Chan}},\ }\href {http://dx.doi.org/10.1103/PhysRevLett.83.1187} {\bibfield
  {journal} {\bibinfo  {journal} {Phys. Rev. Lett.}\ }\textbf {\bibinfo
  {volume} {83}},\ \bibinfo {pages} {1187} (\bibinfo {year}
  {1999})}\BibitemShut {NoStop}%
\bibitem [{\citenamefont {Gasparini}\ \emph {et~al.}(2008)\citenamefont
  {Gasparini}, \citenamefont {Kimball}, \citenamefont {Mooney},\ and\
  \citenamefont {Diaz-Avila}}]{Gasparini2008}%
  \BibitemOpen
  \bibfield  {author} {\bibinfo {author} {\bibfnamefont {F.~M.}\ \bibnamefont
  {Gasparini}}, \bibinfo {author} {\bibfnamefont {M.~O.}\ \bibnamefont
  {Kimball}}, \bibinfo {author} {\bibfnamefont {K.~P.}\ \bibnamefont {Mooney}},
  \ and\ \bibinfo {author} {\bibfnamefont {M.}~\bibnamefont {Diaz-Avila}},\
  }\href {http://link.aps.org/doi/10.1103/RevModPhys.80.1009} {\bibfield
  {journal} {\bibinfo  {journal} {Rev. Mod. Phys.}\ }\textbf {\bibinfo {volume}
  {80}},\ \bibinfo {pages} {1009} (\bibinfo {year} {2008})}\BibitemShut
  {NoStop}%
\bibitem [{\citenamefont {Perron}\ \emph {et~al.}(2010)\citenamefont {Perron},
  \citenamefont {Kimball}, \citenamefont {Mooney},\ and\ \citenamefont
  {Gasparini}}]{Perron2010}%
  \BibitemOpen
  \bibfield  {author} {\bibinfo {author} {\bibfnamefont {J.~K.}\ \bibnamefont
  {Perron}}, \bibinfo {author} {\bibfnamefont {M.~O.}\ \bibnamefont {Kimball}},
  \bibinfo {author} {\bibfnamefont {K.~P.}\ \bibnamefont {Mooney}}, \ and\
  \bibinfo {author} {\bibfnamefont {F.~M.}\ \bibnamefont {Gasparini}},\ }\href
  {http://www.nature.com/nphys/journal/v6/n7/full/nphys1671.html} {\bibfield
  {journal} {\bibinfo  {journal} {Nat. Phys.}\ }\textbf {\bibinfo {volume}
  {6}},\ \bibinfo {pages} {499} (\bibinfo {year} {2010})}\BibitemShut {NoStop}%
\bibitem [{\citenamefont {Perron}\ and\ \citenamefont
  {Gasparini}(2012)}]{Perron2012}%
  \BibitemOpen
  \bibfield  {author} {\bibinfo {author} {\bibfnamefont {J.~K.}\ \bibnamefont
  {Perron}}\ and\ \bibinfo {author} {\bibfnamefont {F.~M.}\ \bibnamefont
  {Gasparini}},\ }\href
  {http://link.aps.org/doi/10.1103/PhysRevLett.109.035302} {\bibfield
  {journal} {\bibinfo  {journal} {Phys. Rev. Lett.}\ }\textbf {\bibinfo
  {volume} {109}},\ \bibinfo {pages} {035302} (\bibinfo {year}
  {2012})}\BibitemShut {NoStop}%
\bibitem [{\citenamefont {Levitin}\ \emph {et~al.}(2007)\citenamefont
  {Levitin}, \citenamefont {Bennett}, \citenamefont {Casey}, \citenamefont
  {Cowan}, \citenamefont {Lusher}, \citenamefont {Saunders}, \citenamefont
  {Drung},\ and\ \citenamefont {Schurig}}]{Levitin2007}%
  \BibitemOpen
  \bibfield  {author} {\bibinfo {author} {\bibfnamefont {L.~V.}\ \bibnamefont
  {Levitin}}, \bibinfo {author} {\bibfnamefont {R.~G.}\ \bibnamefont
  {Bennett}}, \bibinfo {author} {\bibfnamefont {A.}~\bibnamefont {Casey}},
  \bibinfo {author} {\bibfnamefont {B.~P.}\ \bibnamefont {Cowan}}, \bibinfo
  {author} {\bibfnamefont {C.~P.}\ \bibnamefont {Lusher}}, \bibinfo {author}
  {\bibfnamefont {J.}~\bibnamefont {Saunders}}, \bibinfo {author}
  {\bibfnamefont {D.}~\bibnamefont {Drung}}, \ and\ \bibinfo {author}
  {\bibfnamefont {T.}~\bibnamefont {Schurig}},\ }\href
  {http://dx.doi.org/10.1063/1.2828113} {\bibfield  {journal} {\bibinfo
  {journal} {Appl. Phys. Lett.}\ }\textbf {\bibinfo {volume} {91}},\ \bibinfo
  {pages} {262507} (\bibinfo {year} {2007})}\BibitemShut {NoStop}%
\bibitem [{\citenamefont {Mehta}\ and\ \citenamefont
  {Gasparini}(1998)}]{Mehta1998}%
  \BibitemOpen
  \bibfield  {author} {\bibinfo {author} {\bibfnamefont {S.}~\bibnamefont
  {Mehta}}\ and\ \bibinfo {author} {\bibfnamefont {F.~M.}\ \bibnamefont
  {Gasparini}},\ }\href
  {http://link.springer.com/article/10.1023%2FA%3A1022576514910} {\bibfield
  {journal} {\bibinfo  {journal} {J. Low Temp. Phys.}\ }\textbf {\bibinfo
  {volume} {110}},\ \bibinfo {pages} {287} (\bibinfo {year}
  {1998})}\BibitemShut {NoStop}%
\bibitem [{\citenamefont {DeLorenzo}\ and\ \citenamefont
  {Schawb}(2013)}]{DeLorenzo2013}%
  \BibitemOpen
  \bibfield  {author} {\bibinfo {author} {\bibfnamefont {L.~A.}\ \bibnamefont
  {DeLorenzo}}\ and\ \bibinfo {author} {\bibfnamefont {K.~C.}\ \bibnamefont
  {Schawb}},\ }\href {http://arxiv-web3.library.cornell.edu/abs/1308.2164}
  {\bibfield  {journal} {\bibinfo  {journal} {arXiv:1308.2164}\ } (\bibinfo
  {year} {2013})}\BibitemShut {NoStop}%
\bibitem [{\citenamefont {Gonzalez}\ \emph
  {et~al.}(2013{\natexlab{a}})\citenamefont {Gonzalez}, \citenamefont {Zheng},
  \citenamefont {Garcell}, \citenamefont {Lee},\ and\ \citenamefont
  {Chan}}]{Gonzalez2013}%
  \BibitemOpen
  \bibfield  {author} {\bibinfo {author} {\bibfnamefont {M.}~\bibnamefont
  {Gonzalez}}, \bibinfo {author} {\bibfnamefont {P.}~\bibnamefont {Zheng}},
  \bibinfo {author} {\bibfnamefont {E.}~\bibnamefont {Garcell}}, \bibinfo
  {author} {\bibfnamefont {Y.}~\bibnamefont {Lee}}, \ and\ \bibinfo {author}
  {\bibfnamefont {H.~B.}\ \bibnamefont {Chan}},\ }\href
  {http://dx.doi.org/10.1063/1.4790196} {\bibfield  {journal} {\bibinfo
  {journal} {Rev. Sci. Instrum.}\ }\textbf {\bibinfo {volume} {84}},\ \bibinfo
  {pages} {025003} (\bibinfo {year} {2013}{\natexlab{a}})}\BibitemShut
  {NoStop}%
\bibitem [{\citenamefont {Gonzalez}\ \emph
  {et~al.}(2013{\natexlab{b}})\citenamefont {Gonzalez}, \citenamefont {Zheng},
  \citenamefont {Moon}, \citenamefont {Garcell}, \citenamefont {Lee},\ and\
  \citenamefont {Chan}}]{Gonzalez2013b}%
  \BibitemOpen
  \bibfield  {author} {\bibinfo {author} {\bibfnamefont {M.}~\bibnamefont
  {Gonzalez}}, \bibinfo {author} {\bibfnamefont {P.}~\bibnamefont {Zheng}},
  \bibinfo {author} {\bibfnamefont {B.~H.}\ \bibnamefont {Moon}}, \bibinfo
  {author} {\bibfnamefont {E.}~\bibnamefont {Garcell}}, \bibinfo {author}
  {\bibfnamefont {Y.}~\bibnamefont {Lee}}, \ and\ \bibinfo {author}
  {\bibfnamefont {H.~B.}\ \bibnamefont {Chan}},\ }\href
  {http://link.springer.com/article/10.1007%2Fs10909-012-0682-8} {\bibfield
  {journal} {\bibinfo  {journal} {J. Low Temp. Phys.}\ }\textbf {\bibinfo
  {volume} {171}},\ \bibinfo {pages} {200} (\bibinfo {year}
  {2013}{\natexlab{b}})}\BibitemShut {NoStop}%
\bibitem [{\citenamefont {Goldner}\ and\ \citenamefont
  {Ahlers}(1992)}]{Goldner1992}%
  \BibitemOpen
  \bibfield  {author} {\bibinfo {author} {\bibfnamefont {L.~S.}\ \bibnamefont
  {Goldner}}\ and\ \bibinfo {author} {\bibfnamefont {G.}~\bibnamefont
  {Ahlers}},\ }\href {http://link.aps.org/doi/10.1103/PhysRevB.45.13129}
  {\bibfield  {journal} {\bibinfo  {journal} {Phys. Rev. B}\ }\textbf {\bibinfo
  {volume} {45}},\ \bibinfo {pages} {13129} (\bibinfo {year}
  {1992})}\BibitemShut {NoStop}%
\bibitem [{\citenamefont {Hendry}\ and\ \citenamefont
  {McClintock}(1987)}]{Hendry1987}%
  \BibitemOpen
  \bibfield  {author} {\bibinfo {author} {\bibfnamefont {P.~C.}\ \bibnamefont
  {Hendry}}\ and\ \bibinfo {author} {\bibfnamefont {P.~V.~E.}\ \bibnamefont
  {McClintock}},\ }\href {http://dx.doi.org/10.1016/0011-2275(87)90069-5}
  {\bibfield  {journal} {\bibinfo  {journal} {Cryogenics}\ }\textbf {\bibinfo
  {volume} {27}},\ \bibinfo {pages} {131} (\bibinfo {year} {1987})}\BibitemShut
  {NoStop}%
\bibitem [{\citenamefont {Mooney}\ and\ \citenamefont
  {Gasparini}(2002)}]{Mooney2002}%
  \BibitemOpen
  \bibfield  {author} {\bibinfo {author} {\bibfnamefont {K.~P.}\ \bibnamefont
  {Mooney}}\ and\ \bibinfo {author} {\bibfnamefont {F.~M.}\ \bibnamefont
  {Gasparini}},\ }\href
  {http://link.springer.com/article/10.1023%2FA%3A1013720412934} {\bibfield
  {journal} {\bibinfo  {journal} {J. Low Temp. Phys.}\ }\textbf {\bibinfo
  {volume} {126}},\ \bibinfo {pages} {247} (\bibinfo {year}
  {2002})}\BibitemShut {NoStop}%
\bibitem [{\citenamefont {Fisher}\ and\ \citenamefont
  {Barber}(1972)}]{Fisher1972}%
  \BibitemOpen
  \bibfield  {author} {\bibinfo {author} {\bibfnamefont {M.~E.}\ \bibnamefont
  {Fisher}}\ and\ \bibinfo {author} {\bibfnamefont {M.~N.}\ \bibnamefont
  {Barber}},\ }\href {http://link.aps.org/doi/10.1103/PhysRevLett.28.1516}
  {\bibfield  {journal} {\bibinfo  {journal} {Phys. Rev. Lett.}\ }\textbf
  {\bibinfo {volume} {28}},\ \bibinfo {pages} {1516} (\bibinfo {year}
  {1972})}\BibitemShut {NoStop}%
\bibitem [{\citenamefont {Lambert}\ \emph {et~al.}(1980)\citenamefont
  {Lambert}, \citenamefont {Perzinski},\ and\ \citenamefont
  {Salin}}]{Lambert1980}%
  \BibitemOpen
  \bibfield  {author} {\bibinfo {author} {\bibfnamefont {B.}~\bibnamefont
  {Lambert}}, \bibinfo {author} {\bibfnamefont {R.}~\bibnamefont {Perzinski}},
  \ and\ \bibinfo {author} {\bibfnamefont {D.}~\bibnamefont {Salin}},\ }\href
  {http://dx.doi.org/10.1051/jphyslet:0198000410101900} {\bibfield  {journal}
  {\bibinfo  {journal} {J. Physique Lett.}\ }\textbf {\bibinfo {volume} {41}},\
  \bibinfo {pages} {19} (\bibinfo {year} {1980})}\BibitemShut {NoStop}%
\bibitem [{\citenamefont {Donnelly}\ and\ \citenamefont
  {Barenghi}(1998)}]{Donnelly1998}%
  \BibitemOpen
  \bibfield  {author} {\bibinfo {author} {\bibfnamefont {R.~J.}\ \bibnamefont
  {Donnelly}}\ and\ \bibinfo {author} {\bibfnamefont {C.~F.}\ \bibnamefont
  {Barenghi}},\ }\href {http://dx.doi.org/10.1063/1.556028} {\bibfield
  {journal} {\bibinfo  {journal} {J. Phys. Chem. Ref. Data}\ }\textbf {\bibinfo
  {volume} {27}},\ \bibinfo {pages} {1217} (\bibinfo {year}
  {1998})}\BibitemShut {NoStop}%
\bibitem [{\citenamefont {Wilks}(1967)}]{Wilks1967}%
  \BibitemOpen
  \bibfield  {author} {\bibinfo {author} {\bibfnamefont {J.}~\bibnamefont
  {Wilks}},\ }\href@noop {} {\emph {\bibinfo {title} {The Properties of Liquid
  and Solid Helium}}},\ edited by\ \bibinfo {editor} {\bibfnamefont
  {C.}~\bibnamefont {Press}}\ (\bibinfo  {publisher} {Oxford},\ \bibinfo {year}
  {1967})\BibitemShut {NoStop}%
\bibitem [{\citenamefont {Lipa}\ \emph {et~al.}(2003)\citenamefont {Lipa},
  \citenamefont {Nissen}, \citenamefont {Stricker}, \citenamefont {Swanson},\
  and\ \citenamefont {Chui}}]{Lipa2003}%
  \BibitemOpen
  \bibfield  {author} {\bibinfo {author} {\bibfnamefont {J.~A.}\ \bibnamefont
  {Lipa}}, \bibinfo {author} {\bibfnamefont {J.~A.}\ \bibnamefont {Nissen}},
  \bibinfo {author} {\bibfnamefont {D.~A.}\ \bibnamefont {Stricker}}, \bibinfo
  {author} {\bibfnamefont {D.~R.}\ \bibnamefont {Swanson}}, \ and\ \bibinfo
  {author} {\bibfnamefont {T.~C.~P.}\ \bibnamefont {Chui}},\ }\href
  {http://link.aps.org/doi/10.1103/PhysRevB.68.174518} {\bibfield  {journal}
  {\bibinfo  {journal} {Phys. Rev. B}\ }\textbf {\bibinfo {volume} {68}},\
  \bibinfo {pages} {174518} (\bibinfo {year} {2003})}\BibitemShut {NoStop}%
\bibitem [{\citenamefont {Nacher}\ and\ \citenamefont
  {Dupont-Roc}(1991)}]{Nacher1991}%
  \BibitemOpen
  \bibfield  {author} {\bibinfo {author} {\bibfnamefont {P.~J.}\ \bibnamefont
  {Nacher}}\ and\ \bibinfo {author} {\bibfnamefont {J.}~\bibnamefont
  {Dupont-Roc}},\ }\href {http://link.aps.org/doi/10.1103/PhysRevLett.67.2966}
  {\bibfield  {journal} {\bibinfo  {journal} {Phys. Rev. Lett.}\ }\textbf
  {\bibinfo {volume} {67}},\ \bibinfo {pages} {2966} (\bibinfo {year}
  {1991})}\BibitemShut {NoStop}%
\bibitem [{\citenamefont {Maynard}(1976)}]{Maynard1976}%
  \BibitemOpen
  \bibfield  {author} {\bibinfo {author} {\bibfnamefont {J.}~\bibnamefont
  {Maynard}},\ }\href {http://link.aps.org/doi/10.1103/PhysRevB.14.3868}
  {\bibfield  {journal} {\bibinfo  {journal} {Phys. Rev. B}\ }\textbf {\bibinfo
  {volume} {14}},\ \bibinfo {pages} {3868} (\bibinfo {year}
  {1976})}\BibitemShut {NoStop}%
\bibitem [{\citenamefont {Maynard}(1980)}]{Maynard1980}%
  \BibitemOpen
  \bibfield  {author} {\bibinfo {author} {\bibfnamefont {J.~D.}\ \bibnamefont
  {Maynard}},\ }\href {http://dx.doi.org/10.1080/00986448008912530} {\bibfield
  {journal} {\bibinfo  {journal} {Chem. Eng. Commun.}\ }\textbf {\bibinfo
  {volume} {6}},\ \bibinfo {pages} {191} (\bibinfo {year} {1980})}\BibitemShut
  {NoStop}%
\bibitem [{\citenamefont {Barmatz}\ and\ \citenamefont
  {Rudnick}(1968)}]{Barmatz1968}%
  \BibitemOpen
  \bibfield  {author} {\bibinfo {author} {\bibfnamefont {M.}~\bibnamefont
  {Barmatz}}\ and\ \bibinfo {author} {\bibfnamefont {I.}~\bibnamefont
  {Rudnick}},\ }\href {http://link.aps.org/doi/10.1103/PhysRev.170.224}
  {\bibfield  {journal} {\bibinfo  {journal} {Phys. Rev.}\ }\textbf {\bibinfo
  {volume} {170}},\ \bibinfo {pages} {224} (\bibinfo {year}
  {1968})}\BibitemShut {NoStop}%
\bibitem [{\citenamefont {Lambert}\ \emph
  {et~al.}(1979{\natexlab{a}})\citenamefont {Lambert}, \citenamefont
  {Perzinski},\ and\ \citenamefont {Salin}}]{Lambert1979}%
  \BibitemOpen
  \bibfield  {author} {\bibinfo {author} {\bibfnamefont {B.}~\bibnamefont
  {Lambert}}, \bibinfo {author} {\bibfnamefont {R.}~\bibnamefont {Perzinski}},
  \ and\ \bibinfo {author} {\bibfnamefont {D.}~\bibnamefont {Salin}},\ }\href
  {http://dx.doi.org/10.1103/PhysRevB.20.1025} {\bibfield  {journal} {\bibinfo
  {journal} {Phys. Rev. B}\ }\textbf {\bibinfo {volume} {20}},\ \bibinfo
  {pages} {1025} (\bibinfo {year} {1979}{\natexlab{a}})}\BibitemShut {NoStop}%
\bibitem [{\citenamefont {Ferrell}\ and\ \citenamefont
  {Bhattacharjee}(1980)}]{Ferrell1980}%
  \BibitemOpen
  \bibfield  {author} {\bibinfo {author} {\bibfnamefont {R.~A.}\ \bibnamefont
  {Ferrell}}\ and\ \bibinfo {author} {\bibfnamefont {J.~K.}\ \bibnamefont
  {Bhattacharjee}},\ }\href
  {http://link.aps.org/doi/10.1103/PhysRevLett.44.403} {\bibfield  {journal}
  {\bibinfo  {journal} {Phys. Rev. Lett.}\ }\textbf {\bibinfo {volume} {44}},\
  \bibinfo {pages} {403} (\bibinfo {year} {1980})}\BibitemShut {NoStop}%
\bibitem [{\citenamefont {Ferrell}\ \emph {et~al.}(1987)\citenamefont
  {Ferrell}, \citenamefont {Mirhashem},\ and\ \citenamefont
  {Bhattacharjee}}]{Ferrell1987}%
  \BibitemOpen
  \bibfield  {author} {\bibinfo {author} {\bibfnamefont {R.~A.}\ \bibnamefont
  {Ferrell}}, \bibinfo {author} {\bibfnamefont {B.}~\bibnamefont {Mirhashem}},
  \ and\ \bibinfo {author} {\bibfnamefont {J.~K.}\ \bibnamefont
  {Bhattacharjee}},\ }\href {http://link.aps.org/doi/10.1103/PhysRevB.35.4662}
  {\bibfield  {journal} {\bibinfo  {journal} {Phys. Rev. B}\ }\textbf {\bibinfo
  {volume} {35}},\ \bibinfo {pages} {4662} (\bibinfo {year}
  {1987})}\BibitemShut {NoStop}%
\bibitem [{\citenamefont {Pankert}\ and\ \citenamefont
  {Dohm}(1989{\natexlab{a}})}]{Pankert1989a}%
  \BibitemOpen
  \bibfield  {author} {\bibinfo {author} {\bibfnamefont {J.}~\bibnamefont
  {Pankert}}\ and\ \bibinfo {author} {\bibfnamefont {V.}~\bibnamefont {Dohm}},\
  }\href {http://link.aps.org/doi/10.1103/PhysRevB.40.10842} {\bibfield
  {journal} {\bibinfo  {journal} {Phys. Rev. B}\ }\textbf {\bibinfo {volume}
  {40}},\ \bibinfo {pages} {10842} (\bibinfo {year}
  {1989}{\natexlab{a}})}\BibitemShut {NoStop}%
\bibitem [{\citenamefont {Pankert}\ and\ \citenamefont
  {Dohm}(1989{\natexlab{b}})}]{Pankert1989b}%
  \BibitemOpen
  \bibfield  {author} {\bibinfo {author} {\bibfnamefont {J.}~\bibnamefont
  {Pankert}}\ and\ \bibinfo {author} {\bibfnamefont {V.}~\bibnamefont {Dohm}},\
  }\href {http://link.aps.org/doi/10.1103/PhysRevB.40.10856} {\bibfield
  {journal} {\bibinfo  {journal} {Phys. Rev. B}\ }\textbf {\bibinfo {volume}
  {40}},\ \bibinfo {pages} {10856} (\bibinfo {year}
  {1989}{\natexlab{b}})}\BibitemShut {NoStop}%
\bibitem [{\citenamefont {Lambert}\ \emph
  {et~al.}(1979{\natexlab{b}})\citenamefont {Lambert}, \citenamefont {Levelut},
  \citenamefont {Perzynski},\ and\ \citenamefont {Salin}}]{Lambert1979b}%
  \BibitemOpen
  \bibfield  {author} {\bibinfo {author} {\bibfnamefont {B.}~\bibnamefont
  {Lambert}}, \bibinfo {author} {\bibfnamefont {A.}~\bibnamefont {Levelut}},
  \bibinfo {author} {\bibfnamefont {R.}~\bibnamefont {Perzynski}}, \ and\
  \bibinfo {author} {\bibfnamefont {D.}~\bibnamefont {Salin}},\ }\href
  {http://link.springer.com/article/10.1007/BF00113878} {\bibfield  {journal}
  {\bibinfo  {journal} {J. Low Temp. Phys.}\ }\textbf {\bibinfo {volume}
  {37}},\ \bibinfo {pages} {679} (\bibinfo {year}
  {1979}{\natexlab{b}})}\BibitemShut {NoStop}%
\bibitem [{\citenamefont {Lea}\ \emph {et~al.}(1989)\citenamefont {Lea},
  \citenamefont {Spencer},\ and\ \citenamefont {Fozooni}}]{Lea1989}%
  \BibitemOpen
  \bibfield  {author} {\bibinfo {author} {\bibfnamefont {M.~J.}\ \bibnamefont
  {Lea}}, \bibinfo {author} {\bibfnamefont {D.~S.}\ \bibnamefont {Spencer}}, \
  and\ \bibinfo {author} {\bibfnamefont {P.}~\bibnamefont {Fozooni}},\ }\href
  {http://dx.doi.org/10.1103/PhysRevB.39.6527} {\bibfield  {journal} {\bibinfo
  {journal} {Phys. Rev. B}\ }\textbf {\bibinfo {volume} {39}},\ \bibinfo
  {pages} {6527} (\bibinfo {year} {1989})}\BibitemShut {NoStop}%
\bibitem [{\citenamefont {Bhattacharyya}\ and\ \citenamefont
  {Bhattacharjee}(1998)}]{Bhattacharyya1998}%
  \BibitemOpen
  \bibfield  {author} {\bibinfo {author} {\bibfnamefont {S.}~\bibnamefont
  {Bhattacharyya}}\ and\ \bibinfo {author} {\bibfnamefont {J.~K.}\ \bibnamefont
  {Bhattacharjee}},\ }\href {http://link.aps.org/doi/10.1103/PhysRevB.58.15146}
  {\bibfield  {journal} {\bibinfo  {journal} {Phys. Rev. B}\ }\textbf {\bibinfo
  {volume} {58}},\ \bibinfo {pages} {15146} (\bibinfo {year}
  {1998})}\BibitemShut {NoStop}%
\bibitem [{\citenamefont {Rudnick}\ and\ \citenamefont
  {Shapiro}(1965)}]{Rudnick1965}%
  \BibitemOpen
  \bibfield  {author} {\bibinfo {author} {\bibfnamefont {I.}~\bibnamefont
  {Rudnick}}\ and\ \bibinfo {author} {\bibfnamefont {K.~A.}\ \bibnamefont
  {Shapiro}},\ }\href {http://dx.doi.org/10.1103/PhysRevLett.15.386} {\bibfield
   {journal} {\bibinfo  {journal} {Phys. Rev. Lett.}\ }\textbf {\bibinfo
  {volume} {15}},\ \bibinfo {pages} {386} (\bibinfo {year} {1965})}\BibitemShut
  {NoStop}%
\bibitem [{\citenamefont {Pobell}(2006)}]{Pobell2006}%
  \BibitemOpen
  \bibfield  {author} {\bibinfo {author} {\bibfnamefont {F.}~\bibnamefont
  {Pobell}},\ }\href@noop {} {\emph {\bibinfo {title} {Matter and Methods at
  Low Temperatures}}}\ (\bibinfo  {publisher} {Springer},\ \bibinfo {year}
  {2006})\BibitemShut {NoStop}%
\bibitem [{\citenamefont {Aoki}\ \emph {et~al.}(2005)\citenamefont {Aoki},
  \citenamefont {Wada}, \citenamefont {Saitoh}, \citenamefont {Nomura},
  \citenamefont {Okuda}, \citenamefont {Nagato}, \citenamefont {Yamamoto},
  \citenamefont {Higashitani},\ and\ \citenamefont {Nagai}}]{Aoki2005}%
  \BibitemOpen
  \bibfield  {author} {\bibinfo {author} {\bibfnamefont {Y.}~\bibnamefont
  {Aoki}}, \bibinfo {author} {\bibfnamefont {Y.}~\bibnamefont {Wada}}, \bibinfo
  {author} {\bibfnamefont {M.}~\bibnamefont {Saitoh}}, \bibinfo {author}
  {\bibfnamefont {R.}~\bibnamefont {Nomura}}, \bibinfo {author} {\bibfnamefont
  {Y.}~\bibnamefont {Okuda}}, \bibinfo {author} {\bibfnamefont
  {Y.}~\bibnamefont {Nagato}}, \bibinfo {author} {\bibfnamefont
  {M.}~\bibnamefont {Yamamoto}}, \bibinfo {author} {\bibfnamefont
  {S.}~\bibnamefont {Higashitani}}, \ and\ \bibinfo {author} {\bibfnamefont
  {K.}~\bibnamefont {Nagai}},\ }\href
  {http://link.aps.org/doi/10.1103/PhysRevLett.95.075301} {\bibfield  {journal}
  {\bibinfo  {journal} {Phys. Rev. Lett.}\ }\textbf {\bibinfo {volume} {95}},\
  \bibinfo {pages} {075301} (\bibinfo {year} {2005})}\BibitemShut {NoStop}%
\bibitem [{\citenamefont {Davis}\ \emph {et~al.}(2008)\citenamefont {Davis},
  \citenamefont {Pollanen}, \citenamefont {Choi}, \citenamefont {Sauls},\ and\
  \citenamefont {Halperin}}]{Davis2008}%
  \BibitemOpen
  \bibfield  {author} {\bibinfo {author} {\bibfnamefont {J.~P.}\ \bibnamefont
  {Davis}}, \bibinfo {author} {\bibfnamefont {J.}~\bibnamefont {Pollanen}},
  \bibinfo {author} {\bibfnamefont {H.}~\bibnamefont {Choi}}, \bibinfo {author}
  {\bibfnamefont {J.~A.}\ \bibnamefont {Sauls}}, \ and\ \bibinfo {author}
  {\bibfnamefont {W.~P.}\ \bibnamefont {Halperin}},\ }\href
  {http://www.nature.com/nphys/journal/v4/n7/full/nphys969.html} {\bibfield
  {journal} {\bibinfo  {journal} {Nat. Phys.}\ }\textbf {\bibinfo {volume}
  {4}},\ \bibinfo {pages} {571} (\bibinfo {year} {2008})}\BibitemShut {NoStop}%
\bibitem [{\citenamefont {Murakawa}\ \emph {et~al.}(2012)\citenamefont
  {Murakawa}, \citenamefont {Wasai}, \citenamefont {Akiyama}, \citenamefont
  {Wada}, \citenamefont {Tamura}, \citenamefont {Nomura},\ and\ \citenamefont
  {Okuda}}]{Murakawa2012}%
  \BibitemOpen
  \bibfield  {author} {\bibinfo {author} {\bibfnamefont {S.}~\bibnamefont
  {Murakawa}}, \bibinfo {author} {\bibfnamefont {M.}~\bibnamefont {Wasai}},
  \bibinfo {author} {\bibfnamefont {K.}~\bibnamefont {Akiyama}}, \bibinfo
  {author} {\bibfnamefont {Y.}~\bibnamefont {Wada}}, \bibinfo {author}
  {\bibfnamefont {Y.}~\bibnamefont {Tamura}}, \bibinfo {author} {\bibfnamefont
  {R.}~\bibnamefont {Nomura}}, \ and\ \bibinfo {author} {\bibfnamefont
  {Y.}~\bibnamefont {Okuda}},\ }\href
  {http://link.aps.org/doi/10.1103/PhysRevLett.108.025302} {\bibfield
  {journal} {\bibinfo  {journal} {Phys. Rev. Lett.}\ }\textbf {\bibinfo
  {volume} {108}},\ \bibinfo {pages} {025302} (\bibinfo {year}
  {2012})}\BibitemShut {NoStop}%
\bibitem [{\citenamefont {Moon}\ \emph {et~al.}(2010)\citenamefont {Moon},
  \citenamefont {Masuhara}, \citenamefont {Bhupathi}, \citenamefont {Gonzalez},
  \citenamefont {Meisel}, \citenamefont {Lee},\ and\ \citenamefont
  {Mulders}}]{Moon2010}%
  \BibitemOpen
  \bibfield  {author} {\bibinfo {author} {\bibfnamefont {B.~H.}\ \bibnamefont
  {Moon}}, \bibinfo {author} {\bibfnamefont {N.}~\bibnamefont {Masuhara}},
  \bibinfo {author} {\bibfnamefont {P.}~\bibnamefont {Bhupathi}}, \bibinfo
  {author} {\bibfnamefont {M.}~\bibnamefont {Gonzalez}}, \bibinfo {author}
  {\bibfnamefont {M.~W.}\ \bibnamefont {Meisel}}, \bibinfo {author}
  {\bibfnamefont {Y.}~\bibnamefont {Lee}}, \ and\ \bibinfo {author}
  {\bibfnamefont {N.}~\bibnamefont {Mulders}},\ }\href
  {http://link.aps.org/doi/10.1103/PhysRevB.81.134526} {\bibfield  {journal}
  {\bibinfo  {journal} {Phys. Rev. B}\ }\textbf {\bibinfo {volume} {81}},\
  \bibinfo {pages} {134526} (\bibinfo {year} {2010})}\BibitemShut {NoStop}%
\bibitem [{\citenamefont {Cai}\ \emph {et~al.}(2001)\citenamefont {Cai},
  \citenamefont {Liu},\ and\ \citenamefont {Lam}}]{Cai2001}%
  \BibitemOpen
  \bibfield  {author} {\bibinfo {author} {\bibfnamefont {C.}~\bibnamefont
  {Cai}}, \bibinfo {author} {\bibfnamefont {G.~R.}\ \bibnamefont {Liu}}, \ and\
  \bibinfo {author} {\bibfnamefont {K.~Y.}\ \bibnamefont {Lam}},\ }\href
  {http://dx.doi.org/10.1006/jsvi.2001.3775} {\bibfield  {journal} {\bibinfo
  {journal} {J. Sound Vib.}\ }\textbf {\bibinfo {volume} {248}},\ \bibinfo
  {pages} {71} (\bibinfo {year} {2001})}\BibitemShut {NoStop}%
\bibitem [{\citenamefont {Vignos}\ and\ \citenamefont
  {Fairbank}(1966)}]{Vignos1966}%
  \BibitemOpen
  \bibfield  {author} {\bibinfo {author} {\bibfnamefont {J.~H.}\ \bibnamefont
  {Vignos}}\ and\ \bibinfo {author} {\bibfnamefont {H.~A.}\ \bibnamefont
  {Fairbank}},\ }\href {http://link.aps.org/doi/10.1103/PhysRev.147.185}
  {\bibfield  {journal} {\bibinfo  {journal} {Phys. Rev.}\ }\textbf {\bibinfo
  {volume} {147}},\ \bibinfo {pages} {185} (\bibinfo {year}
  {1966})}\BibitemShut {NoStop}%
\end{thebibliography}%

\end{document}